\documentstyle[12pt,preprint,aps,epsf]{revtex}

\newcommand{\qtab}[4]{
\begin{minipage}{0.2\textwidth}
\begin{center} #1\end{center} \end{minipage}& \begin{minipage}{0.26\textwidth}
\begin{center} #2  \end{center} \end{minipage}& \begin{minipage}{0.26\textwidth}
\begin{center} #3  \end{center} \end{minipage}& \begin{minipage}{0.26\textwidth}
\begin{center} #4 \end{center} \end{minipage} \\ \hline
} %  }
\newcommand{\qtabstart}{\begin{tabular}{|c|c|c|c|} \hline }
\newcommand{\qtabstop}{\end{tabular}}

\newcommand{\JKalt}[1]{}

\begin{document}

\draft
\tighten
\title{Modelling the many-body dynamics of heavy ion collisions:
       Present status and future perspective}

\author{Ch.\ Hartnack, Rajeev K.\ Puri, J.\ Aichelin}
\address{Laboratoire de subatomique et des Technologies associ\'ees \\
(SUBATECH), UMR Universit\'e de Nantes, CNRS, Ecole des Mines \\ 
4 rue A. Kastler, F-44070 Nantes, France} 
\author{J.\ Konopka, S.A.\ Bass, H. St\"ocker, W. Greiner} 
\address{Institut f\"ur Theoretische Physik, 
         Johann Wolfgang Goethe-Universit\"at\\
         Postfach 11 19 32, D-60054 Frankfurt am Main, Germany}

\maketitle

\begin{abstract}
Basic problems of the semiclassical
microscopic modelling of strongly interacting
systems are discussed within the framework of Quantum Molecular Dynamics (QMD).
This model allows to study the influence of several types
of nucleonic interactions on a large variety of observables and 
phenomena occurring in heavy ion 
collisions at relativistic energies.
It is shown that the same predictions can be obtained with several --
numerically completely different and independently written -- programs as far
as the same model parameters are employed and the same basic approximations 
are made. Many observables are robust against variations of the details of 
the model assumptions used. Some of the physical results, however, depend 
also on rather technical parameters like the preparation of the initial 
configuration in phase space. This crucial problem is connected with the 
description  of the ground state of single nuclei, 
which differs among the various approaches. 
%For one particular version of the QMD model his effect introduces significant
%systematic errors which are critically examined. 
An outlook to an improved
molecular dynamics scheme for heavy ion collisions is given.
\JKalt{
Numerical realizations of the Quantum Molecular Dynamics (QMD) model are
discussed and compared. It is found, that these programs -- although
numerically completely different and independently written --
yield the same physical output if the
same basic approximations are made. It is shown that most of these differences 
between the results of the employed 
programs can be related to the  
%%% conceptual ansatzes mainly in the 
descriptions of the ground state of the colliding nuclei. 
Other ingredients like different parametrizations of the cross sections
produce only small effects.}
\end{abstract}


%%%%%%%%%%%%
%\psdraft
%%%%%%%%%%%

%%%%%%%%%%%%%%%%%%%%%%%%%%%%%%%%%%%%%%%%%%%%%%%%%%%%%%%%%%%%%%%%%%%%%%%%%%
\section{Introduction}
One of the main interests of the study of relativistic heavy ion collisions 
is the investigation of the properties of nuclear matter at extreme 
densities and excitation energies.
\cite{sch68,cse86,sto86,st86,cl86,schue87,cas90,ai91}.
These investigations include the production of secondary particles, the 
properties of particles in a (dense) nuclear medium, the compression and
repulsion of dense nuclear matter, its equilibration during the reaction and
its decay into fragments and single particles.
On a macroscopic level the total  energy of a dense nuclear system 
and its decomposition into thermal and compressional parts is related to
the concept of the nuclear equation of state.
Since a consistent derivation of the nuclear equation of state, e.g.\ the
energy per nucleon as a function of 
density and temperature, is only possible in the low density limit
(Br\"uckner theory) a reliable theoretical description is not at hand. 
On the other hand this quantity is of interest for many astrophysical 
questions \cite{astro}
and therefore its knowledge is highly desirable. Heavy ion 
reactions in combination with corresponding
simulations using a variety of 
parametrizations of the equation of state
are presently the only possible approach to study this quantity.

Heavy ion collisions allow to search for a large number 
of observables which may be used as indicators of the properties of matter
under extreme conditions. Frequently these observables are related to the
quantitative description of collective effects like  
the bounce--off of cold spectator matter in the reaction plane 
\cite{st80} 
and the squeeze--out of hot and compressed participant matter perpendicular
to the reaction plane 
\cite{st82} 
as well as to the  production of secondary particles
\cite{st78,da79,st81}.

Experiments performed at LBL in the early 80's
(Streamer chamber, Plastic ball) yield
first $4\pi$ information of the final momentum distributions in heavy ion 
reactions 
\cite{flow,pions}.
New experimental $4\pi$ setups at LBL, Ganil, GSI and Brookhaven enable precise  
measurements on the emission of primary and secondary particles and 
therefore provide a stimulating challenge to the theoretical description 
of heavy ion collisions 
%\cite{}. 

Lots of comparisons have been made between experimental data and microscopic
and macroscopic transport-theoretical calculations. 
Besides other microscopic models like VUU 
\cite{kru85a}, 
BUU 
\cite{bert84a,ai85a},
Landau-Vlasov 
\cite{greg87}, 
AMD
\cite{amd} 
or FMD 
\cite{fmd}
the Quantum Molecular Dynamics approach (QMD) 
is a frequently used model
\cite{ai86}. 
However, from recent comparisons of experimental 
results with QMD using different numerical realizations conflicting results 
have been reported \cite{qmdvsqmd}. 
We will demonstrate that these discrepancies
are on the one hand due to the variation of physical parameters (like 
ground state densities, interaction ranges) whose precise values are not  
known. On the other hand they are a consequence of the impossibility to 
build a ground state nucleus with all its detailed structure in a semiclassical 
molecular approach.

This paper is organized as follows: 
First the basic principles of a microscopic modelling of heavy ion reactions 
are briefly reported. The assumptions entering in the different QMD 
realizations are described in detail.
The origin of differences is critically examined in this context.
We demonstrate that most of the 
discrepancies can be attributed to different descriptions of the initial
nuclei, which limits the applicability of some versions. 
Finally an outlook to a new molecular dynamics scheme for 
heavy ion collisions simulations is given. 

%%%%%%%%%%%%%%%%%%%%%%%%%%%%%%%%%%%%%%%%%%%%%%%%%%%%%%%%%%%%%%%%%%%%%%%%%%
\section{Microscopic modelling of heavy ion reactions}
Presently the microscopic models can be subdivided into two classes:
Those which follow the time evolution of the one-body phase space distribution
and those which are based on $n$-body molecular dynamics or cascade 
schemes.

%%%%%%%%%%%%%%%%%%%%%%%%%%%%%%%%%%%%%%%%%%%%%%%%%%%%%%%%%%%%%%%%%%%%%%%%%%
\subsection{VUU-type models}
The microscopic transport models for the one-body Wigner phase space 
density distribution obtained different names although they solve
the same equation. They differ in the technical realization, i.e.\ the 
computer program, and are known as Vlasov--Uehling--Uhlenbeck (VUU) 
model \cite{kru85a,moli85b} (or BUU \cite{bert84a,ai85a}, LV \cite{greg87} ).
They solve the following transport equation  for the one-body Wigner
density $f(\vec{r},\vec{p},t)$ in the limit $\hbar \to 0$:
\JKalt{
\begin{eqnarray}  \label{vuueq}
   \frac{\partial f}{\partial t} &+&
   \vec{v} \cdot \nabla_r f - \nabla_r U \cdot \nabla_p f \quad
 = \quad - \frac{1}{(2\pi)^6} \int {\rm d}^3 p_2 \,{\rm d}^3 p_2 ' \,
 {\rm d}^3 p_1 ' \,
%%  {\rm d}\Omega \,
   \sigma   v_{12} 
%% \frac{{\rm d}\sigma}{{\rm d}\Omega} 
 \nonumber \\
   &\times& \left [ f f_2 (1-f_1 ') (1-f_2 ') -f_1 ' f_2 ' (1-f)
   (1-f_2) \right ]  \nonumber \\
&\times&\delta^4 (p+p_2 -p_1 ' - p_2 ' ).
\end{eqnarray}
}
\begin{eqnarray}  \label{vuueq}
   \frac{\partial f}{\partial t} &+&
   \vec{v} \cdot \nabla_r f - \nabla_r U \cdot \nabla_p f \quad
 = \quad - \frac{4 \pi^3 (\hbar c)^4}{\hbar (mc^2)^2} 
\int \frac{{\rm d}^3 p_1'}{(2\pi \hbar)^3} \,
\frac{{\rm d}^3 p_2 '}{(2\pi \hbar)^3} \,
 {\rm d}^3 p_2 \,
%%  {\rm d}\Omega \,
%%   \sigma   v_{12} 
%% 
\frac{{\rm d}\sigma}{{\rm d}\Omega} 
 \nonumber \\
   &\times& \left [ f f_2 (1-f_1 ') (1-f_2 ') -f_1 ' f_2 ' (1-f)
   (1-f_2) \right ]  \nonumber \\
&\times&\delta^4 (p+p_2 -p_1 ' - p_2 ' ).
\end{eqnarray}

The l.h.s.\ of this equation is the total differential of $f$ 
with respect to the time assuming a momentum independent potential $U$.
This potential is calculated selfconsistently and corresponds to
the real part of the Br\"uckner G-matrix. Usually a Skyrme-parametrization
\begin{equation} \label{skyrmeansatz}
 U \,=\, \alpha \left(\frac{\rho}{\rho_0} \right) + 
	\beta \left( \frac{\rho}{\rho_0} \right)^{\gamma}
\end{equation}
of the real part of the G-matrix is employed, where $\rho$ is the nuclear
density which is frequently measured in units of the saturation density
$\rho_0$ of cold nuclear matter.

The r.h.s.\ of Eq.\ (\ref{vuueq}) contains a Boltzmann collision integral, 
which is identified with the imaginary part of the G-matrix. 
This part describes the influence of binary hard-core collisions, where
the term with $ff_2$ describes the loss of particles (in a phase space region)
and the term with $f_1'f_2'$ the gain term due to collisions feeding the 
considered phase space region. It is supplemented
with the Nordheim--Uehling--Uhlenbeck modifications in order to obey the 
Pauli-principle in the final state of the collisions 
\cite{ue33}.
The $\delta$-functions assure the conservation of the four-momentum.
The cross section $\sigma$
is normally adjusted to the free nucleon-nucleon scattering.
The differences from cross sections calculated from the imaginary part of the
Br\"uckner G-matrix are minor \cite{boh91} and influence little the observables
of a heavy ion collision. For a derivation of this equation see 
\cite{malfliet,mosel}. 

The equation is solved by use of the testparticle method. 
Here the continuous one-body distribution function $f$ at $t=0$ is represented
by an ensemble of $n \cdot(A_p + A_t)$  pointlike particles.
This is often viewed as an ensemble of $n$ parallel events with $A_p+A_t$ 
physical particles each, where $A_p$ and $A_t$ denote
the number of nucleons in projectile and target, respectively. 
The l.h.s.\ of Eq.\ (\ref{vuueq}) can be regarded as the transport equation 
(Vlasov-equation) for a distribution 
of classical particles whose time evolution is governed by 
Hamilton's equations of motion.
\begin{equation}
\label{HamEq3}
\dot{\vec{p}}_i = - \frac{\partial \langle H \rangle}{\partial \vec{r}_i} 
\quad {\rm and} \quad
\dot{\vec{r}}_i = \frac{\partial \langle H \rangle}{\partial \vec{p}_i} \, ,
\end{equation}
The testparticles move due to their own, selfconsistently
generated mean-field. 
The r.h.s. is taken into account by
additional stochastic
scattering similar to the collisions in cascade models\cite{yar79,cug80}.

More explicitly the test particle method corresponds to the replacement of the 
expectation value of a single particle observable 
\begin{equation}
\langle O(t) \rangle = \int f(\vec{r},\vec{p},t) 
O(\vec{r},\vec{p}) \, d^3r\, d^3 p 
\end{equation}
by a Monte Carlo integration
\begin{equation}
\label{MC5}
\langle O(t) \rangle = \frac{1}{n(A_T+A_P)} \sum_{i=1}^{n(A_T+A_P)} 
O(\vec{r}_i(t),\vec{p}_i(t)) 
\end{equation}
where the $\vec{r}_i(t)$ and $\vec{p}_i(t)$ are points in phase 
space which are distributed according to $f(\vec{p},\vec{r},t)$, i.e.,
\begin{equation} \label{fapprox}
f(\vec{p},\vec{r},t) = \lim_{n\rightarrow \infty}
\frac{1}{n(A_T+A_P)} \sum_{i=1}^{n(A_T+A_P)} 
 \delta(\vec{r}-\vec{r}_i(t))\delta(\vec{p}-\vec{p}_i(t))
\end{equation}

It is evident that a large number $n$ is necessary to avoid numerical 
noise. Predictions beyond the one-body level are not feasible although several 
attempts have been made to relate the (unphysical) numerical noise to physical 
fluctuations. %%%%% \cite{suraud}. 
In practice the number $n$ lies in the range between 15 and 500 and one employs
a grid to obtain a smooth phase space density distribution.

The numerical realization can be achieved in various ways. 
VUU uses a phase space sphere around each particle in order to 
determine $f$ and a coordinate space sphere to
determine $\rho$ and thus $U(\rho)$. This corresponds to a Lagrangian
method. On the contrary, BUU uses  a fixed grid corresponding to an 
Eulerian method in hydrodynamics. In both models 
collisions are treated in a parallel event method, only testparticles of the 
same events, i.e. the $A_p + A_t$ test particles with the same index $n$, 
can collide. The Landau-Vlasov model determines $f$ by the overlap
of several Gaussians. The collisions are performed in a crossed-event (or full
ensemble) method where all $n(A_p+A_t)$ 
may collide with each other particle with a scaled 
cross section.

For a solution of Eq.\ (\ref{vuueq}) proper boundary conditions have to
be specified. In the case of heavy ion reactions, the test particles 
are distributed according to the density- and (Fermi-) momentum distribution 
of ground state nuclei. The latter are then boosted onto every other with 
the proper relative momentum.
%The requirement of a proper description of the ground state of the two 
%colliding nuclei we can use as initial condition of the test particles.
%\begin{equation} \label{fwiththeta}
%f(\vec{r},\vec{p},t=0) \approx \frac{1}{n(A_T+A_P)} \sum_{i=1}^{n(A_T+A_P)} 
%\Theta (\vec{r}-\vec{r}_i(t=0))\Theta(\vec{p}-\vec{p}_i(t=0)).
%\end{equation}
%$\Theta(\vec{r})$ should be in principle a Dirac function $\delta(\vec{r})$ 
%as indicated in eq. (\ref{fapprox}).
%Practically it is frequently replaced by a volume sphere (or sometimes by 
%a cubic box)
%\begin{equation} \label{thetadef}
%\Theta(\vec{r}) = \left \{ \begin{array}{ll} 
%3/(4\pi R^3) & |\vec{r}| \le R \\
%0  &  |\vec{r}| > R 
%\end{array} \right.
%\end{equation}
%with $R \to 0$ for $n \to \infty$.
Initially the test particles are randomly distributed in a coordinate space 
sphere of the radius $R= 1.12 A^{1/3} fm$ (where A is the atomic number of 
the nucleus) and in a momentum space sphere of the radius of the corresponding
Fermi momentum.
 
One should keep in mind that the forces acting on the testparticles
are calculated from the entire distribution including testparticles from
all events, hence the $n$ parallel events are not independent and
event-by-event correlations cannot be analyzed within
this one-body transport models.
In the limit $n \to \infty$ the distribution of these propagated test 
particles at the time $t$ represents the one-body distribution function
at this time. Any one-body observable can be calculated by averaging 
the values weighted with the distribution function according to Eq.\ (\ref{MC5}). 
Hence, VUU type models succeeded in the description of one-body
observables like collective flow, stopping and particle spectra, but,
fluctuations and correlations, such as the formation of 
fragments or the description of two-particle correlations
in relativistic heavy ion collisions, are beyond the scope of a 
transport model based on a one--body distribution function 
\cite{gossiaux951,gossiaux952}. 
Any fluctuation of the observables 
seen in the Monte Carlo simulation of the one--body distribution function 
is due to numerical noise and disappears in the limit of a infinite number
of test particles. 

%This was one of the motivations 
%for the development of the Quantum Molecular Dynamics model (QMD) 
%\cite{ai87b,ai86,pei89,ai91}. 

%%%%%%%%%%%%%%%%%%%%%%%%%%%%%%%%%%%%%%%%%%%%%%%%%%%%%%%%%%%%%%%%%%%%%%%%%%
\subsection{The Quantum Molecular Dynamics approach}
An approach which goes beyond a one-body description as explained above,
is the Quantum Molecular Dynamics (QMD) model
\cite{ai87b,ai86,pei89,ai91}. 
The QMD model is a $n$-body theory which simulates heavy ion reactions 
at intermediate energies on an event by event basis. 
Taking into account all fluctuations and correlations has basically two
advantages: i) many-body processes, in particular the formation of complex
fragments are explicitly treated and ii) the model allows for an 
event-by-event analysis of heavy ion reactions similar to the methods which
are used for the analysis of exclusive high acceptance data.

The major aspects of the formulation of QMD will now be discussed briefly.
For a more detailed description we refer to ref.~\cite{ai91}.
The particular realizations of this model will be discussed later.

%%%%%%%%%%%%%%%%%%%%%%%%%%%%%%%%%%%%%%%%%%%%%%%%%%%%%%%%%%%%%%%%%%%%%%%%%%
\subsubsection{Formal derivation of the transport equation}
In QMD each nucleon is represented
by a coherent state of the form (we set $\hbar,c =1$) 
which are characterized by 6 time-dependent parameters,
$\vec{r}_{i}$ and $\vec{p}_{i}$, respectively.
\begin{equation}
\label{gaussians}
%\phi_j(\vec{x}_j; \vec{r}_j(t), \vec{p}_j(t)) =
%\left(\frac{2}{L \pi}\right)^{3/4} {\rm e}^{-(\vec{x}_j-\vec{r}_j(t))^2/L +
%{\rm i} \vec{p}_j(t)\vec{x}_j)}
\phi_i (\vec{x_i};t) = 
\left({\frac{2 }{L\pi}}\right)^{3/4}\, e^{-(\vec{x_i} -
\vec{r_i}(t) %%% -\vec{p_i}(t) t/m
)^2/L} \,e^{i(\vec{x_i} %%%-
%%%\vec{r_i}(t)) 
\vec{p_i}(t)} %\,
% e^{-i p_i^2(t)t/2m}
. 
\end{equation}
The parameter $L$, which is related  to the extension of the wave packet in
phase space, is fixed. 
The total $n$-body wave function is  assumed to be the direct
product of coherent states (\ref{gaussians})
\begin{equation}
%\Phi = \prod_{j=1}^n \phi_j(\vec{x}_j)
\Phi = \prod_i \phi_i (\vec{x_i}, \vec{r_i}, \vec{p_i}, t)  
\end{equation}
Note that we do not use a Slater determinant 
(with $(A_p+A_t)$! summation terms) and
thus neglect antisymmetrization. First successful attempts 
to simulate heavy ion reactions with antisymmetrized
states have been performed for small systems \cite{fmd,amd}. 
A consistent derivation of the QMD equations of motion for the wave function
under the influence of both, the real and the imaginary part of the 
G-matrix, however, has not yet been achieved. Therefore we will add the 
imaginary part as a cross section and treat them as in the cascade approach.
How to incorporate cross sections into a antisymmetrized molecular 
dynamics is not yet known. This limits its applicability to very low beam 
energies.

The initial values of the parameters
are chosen in a way that the ensemble of $A_T$ + $A_P$ 
nucleons gives a proper density distribution as well as a proper momentum
distribution of the projectile and target nuclei. 

The equations of motion of the many-body 
system is calculated by means of a generalized variational principle: we
start out from the action \cite{KermannKoonin}
\begin{equation}
S = \int\limits ^{t_2} _{t_1} {\cal L} [\Phi, \Phi^\ast] dt 
\end{equation}
with the Lagrange functional ${\cal L}$ 
\begin{equation}
{\cal L} = \left \langle\Phi \left\vert i\hbar {\frac{d }{dt}} - H \right\vert
\Phi\right\rangle
\end{equation}
where the total time derivative includes the derivation with respect to the
parameters. 
The Hamiltonian $H$ contains a kinetic term and mutual 
interactions $V_{ij}$, which can be interpreted as the real part of the 
Br\"uckner G-matrix supplemented by the Coulomb interaction. We will lateron 
describe the components of $H$ in detail.
The time evolution of the parameters is obtained by the
requirement that the action is stationary under the allowed variation of the
wave function. This yields an Euler-Lagrange equation for each parameter.

If the true solution of the Schr\"odinger equation is contained in the
restricted set of wave functions 
$\phi_i(\vec{x}_i, t)$ (with parameters $\vec{r_i},\vec{p_i}$) 
%%%%% $\phi_\alpha(x_1,x_\alpha,p_\alpha)$ 
this variation of the action gives the exact solution of the Schr\"odinger
equation. If the parameter space is too restricted we obtain that wave
function in the restricted parameter space which comes closest to the
solution of the Schr\"odinger equation. Note that the set of wave functions
which can be covered with special parametrizations is not necessarily a
subspace of Hilbert-space, thus the superposition principle does not hold.

For the coherent states and a Hamiltonian of the form $H = \sum_i T_i + {\ 
\frac{1 }{2}} \sum_{ij} V_{ij}$ ($T_i$= kinetic energy, $V_{ij}$ = potential
energy) the Lagrangian and the variation can easily be calculated and we
obtain: 
\begin{equation}
{\cal L} = \sum_i \biggl[-\dot{\vec{r}_i} {\vec{p}_i}  - T_i -
{1\over 2}\sum _{j\neq i} \langle
V_{ik}\rangle - {\frac{3 }{2Lm}} \biggr].\end{equation}
Variation yields:
\begin{equation}
\dot{\vec{r}_i} = {\frac{\vec{p}_i }{m}} + \nabla_{\vec{p}_i} \sum_j
\langle V_{ij}\rangle  = \nabla_{\vec{p}_i} \langle H \rangle
\end{equation}
\begin{equation}
\dot{\vec{p}_i} = - \nabla_{\vec{r}_i} \sum _{j\neq i} \langle
V_{ij}\rangle = -\nabla_{\vec{r}_i} \langle H \rangle
\end{equation}
with $\langle V_{ij}\rangle = \int d^3x_1\,d^3x_2\, 
\phi_i^* \phi_j^* V(x_1,x_2) \phi_i \phi_j$.
These are the time evolution equations which are solved numerically. 
Thus the variational principle reduces the time evolution of the 
$n$-body Schr\"odinger equation to the time evolution equations of 
$6 \cdot (A_P+A_T)$ parameters to which a physical meaning can be
attributed.
The equations of motion for the parameters $\vec{p}_i$ and $\vec{r}_i$ read
\begin{equation}\label{hamiltoneq}
\dot{\vec{p}}_i = - \frac{\partial \langle H \rangle}{\partial \vec{r}_i} 
\quad {\rm and} \quad
\dot{\vec{r}}_i = \frac{\partial \langle H \rangle}{\partial \vec{p}_i} \, ,
\end{equation}
and show the same structure as the classical Hamilton equations, Eq.\ 
(\ref{HamEq3}).
The numerical solution can be treated in a similar manner as it is done in
classical molecular dynamics 
\cite{bod77,wil78,kis83,mol84}.
% and corresponds to the solution of a
%classical Liouville equation. In contradistinction to the transport equation
%for the one-body phase space density QMD conserves energy and momentum by
%construction.
Trial wave functions other than the gaussians in Eq.\ (\ref{gaussians}), yield
more complex equations of motion for other parameters and hence the analogy 
to classical molecular dynamics is lost. If $\langle H \rangle$ has no
explicit time dependence, QMD conserves energy and momentum by construction.

%However, the G - matrix becomes complex if the energy is sufficiently high.
%In this case the imaginary acts like a cross section 
%(for details see \cite{ai91}). 
%This forces the inclusion of additional quantum features like
%stochastic scattering (which has to respect Pauli's principle) 
%and particle production.  The inclusion of the collisions is done
%similar as  the Boltzmann collision term with Uehling-Uhlenbeck
%modifications of VUU. We will soon come back to this point. 

%%%%%%%%%%%%%%%%%%%%%%%%%%%%%%%%%%%%%%%%%%%%%%%%%%%%%%%%%%%%%%%%%%%%%%%%%%
\subsubsection{Description of the Hamiltonian}
The nuclear dynamics of the QMD can also be translated into a
semiclassical scheme. The Wigner distribution function $f_i$ of 
the nucleon $i$ can be easily derived 
from the test wave functions (note that antisymmetrization is neglected). 
\begin{equation} \label{fdefinition}
 f_i (\vec{r},\vec{p},t) = \frac{1}{\pi^3 \hbar^3 }
 {\rm e}^{-(\vec{r} - \vec{r}_{i} (t) )^2  \frac{2}{L} }
 {\rm e}^{-(\vec{p} - \vec{p}_{i} (t) )^2  \frac{L}{2\hbar^2}  }
\end{equation} 
and the total Wigner density is the sum of those of all nucleons.
Hence the expectation value of the total Hamiltonian reads 
\begin{eqnarray} 
\langle H \rangle &=& \langle T \rangle + \langle V \rangle 
\nonumber \\ \label{hamiltdef}
&=& \sum_i \frac{p_i^2}{2m_i} +
\sum_{i} \sum_{j>i}
 \int f_i(\vec{r},\vec{p},t) \,
V^{ij}  f_j(\vec{r}\,',\vec{p}\,',t)\,
d\vec{r}\, d\vec{r}\,'
d\vec{p}\, d\vec{p}\,' \quad.
\end{eqnarray}
The baryon-potential consists of the real part of the 
$G$-Matrix which is supplemented by the Coulomb interaction
between the charged particles. The former can be further subdivided in 
a part containing the contact Skyrme-type interaction only, a contribution
due to a finite range Yukawa-potential, and a momentum dependent part.
$V^{ij} = G^{ij}+V^{ij}_{Yuk}+V^{ij}_{Coul}+ V^{ij}_{mdi} $
 %%%%%+V^{ij}_{sym} $
consists of 
\begin{eqnarray}
V^{ij} &=& G^{ij} + V^{ij}_{\rm Coul} \nonumber \\
       &=& V^{ij}_{\rm Skyrme} + V^{ij}_{\rm Yuk} + V^{ij}_{\rm mdi} + 
           V^{ij}_{\rm Coul} \nonumber \\
       &=& t_1 \delta (\vec{x}_i - \vec{x}_j) + 
           t_2 \delta (\vec{x}_i - \vec{x}_j) \rho^{\gamma-1}(\vec{x}_{i}) +
           t_3 \frac{\hbox{exp}\{-|\vec{x}_i-\vec{x}_j|/\mu\}}
               {|\vec{x}_i-\vec{x}_j|/\mu} + \label{vijdef}  \\
       & & t_4\hbox{ln}^2 (1+t_5(\vec{p}_i-\vec{p}_j)^2)
               \delta (\vec{x}_i -\vec{x}_j) +
           \frac{Z_i Z_j e^2}{|\vec{x}_i-\vec{x}_j|} \nonumber
\end{eqnarray}
$Z_i,Z_j$ are the charges of the baryons $i$ and $j$. 
The real part of the Br\"uckner G-matrix is density dependent, which is
reflected in the expression for $G^{ij}$.
The expectation value of $G$ for the nucleon $i$ 
is a function of the interaction density 
$\rho_{\rm int}^i$. It is indeed this quantity which relates the number density
to the energy content of nuclear matter.
\begin{equation} \label{rhoint}
\rho_{\rm int}^i(\vec{r_i}) = \frac{1}{(\pi L)^{3/2}} 
\sum_{j \neq i} {\rm e}^{\displaystyle 
-(\vec{r_{i}}-\vec{r_{j}})^2/L }
\end{equation}
Note that the interaction density 
has twice the width of the single particle density. Moreover,
the particles do not interact with themselves. This is different compared to
VUU-type models because in QMD 
explicit $N$--$N$ interactions are treated, hence the force acting
on a particle at the position $\vec{r}$ depends on the exact positions of 
all other particles, whereas the density employed in the one-body theories
(eq. (\ref{fapprox})) 
depends on the average number of nucleons in the vicinity of the
test particle only. 

It should be noted that the width $L$ of the distribution function determines
the interaction range of the particle and influences the density distribution
of finite systems. Therefore its value has to be adopted to reasonable 
interaction ranges of the strong interaction. 

The momentum dependence $V_{\rm mdi}^{ij}$ of the $N$--$N$ interaction,
which may optionally be used in QMD, is fitted to
experimental data \cite{ar82,pa67} on
the real part of the nucleon optical potential \cite{schue87,ai87b,bert88b},
which yields
\begin{equation}
\label{mdipar}
U_{mdi} =        \delta \cdot \mbox{ln}^2 \left( \varepsilon \cdot  
                \left( \Delta \vec{p} \right)^2 +1 \right) \cdot
                        \left(\frac{\rho_{int}}{\rho_0}\right)
\end{equation}
These measurements have been superseded recently by new data
\cite{hama90} and thus a new parametrization has been advanced \cite{newopt}.

The potential part of the equation of state (we will discuss this concept 
in the next subsection in more detail) 
resulting from the convolution of 
the distribution
functions $f_i$ and $f_j$ with the interactions  
$V_{\rm Skyrme}^{ij}+ V_{\rm mdi}^{i,j}$ 
(local interactions including
momentum dependence) then reads:
\begin{equation} \label{eosinf}
U \,=\, \alpha \cdot \left(\frac{\rho_{int}}{\rho_0}\right) +
        \beta \cdot \left(\frac{\rho_{int}}{\rho_0}\right)^{\gamma} +
        \delta \cdot \mbox{ln}^2 \left( \varepsilon \cdot
                \left( \Delta \vec{p} \right)^2 +1 \right) \cdot
                        \left(\frac{\rho_{int}}{\rho_0}\right)
\end{equation}

Here it should be noted that due to the definition of $\rho_{int}$ 
(eq. \ref{rhoint})
no mean-field potentials (as e.g. eq. (\ref{skyrmeansatz}) for VUU) 
show up in the calculation of the equations of 
motion (eq. \ref{hamiltoneq} ) of QMD but a sum of two (and three) body
interactions (see eq. \ref{vijdef}). 
Hence energy and momentum  are - in contrast to  single VUU `events'- 
strictly conserved in each event.

The Coulomb interaction cannot be treated for infinite matter, since this
leads to diverging terms. In the first versions of QMD no explicit
treatment of the isospin is 
performed and the charges are replaced by effective
charges, i.e.\ all nucleons had been attributed the effective charge
$Z = (Z_{proj.}+Z_{targ.})/(A_{proj.} +A_{targ.})$. IQMD (we will
later come to that) and other more recent versions use the real baryon charges.
  
The parameters $\mu$ and $t_1 ... t_5$ are adjusted to fit the real
part of the G-matrix and to describe the properties of finite
nuclei.

%%%%%%%%%%%%%%%%%%%%%%%%%%%%%%%%%%%%%%%%%%%%%%%%%%%%%%%%%%%%%%%%%%%%%%%%%%
\subsubsection{The relation to the nuclear equation of state}
One strong motivation for the numerical simulation of heavy ion
reactions is the possibility to investigate effects of the underlying
nuclear equation of state on the dynamics and final states of these
collisions. QMD is a model for non-equilibrium dynamics
with mutual interactions among the constituents and therefore does not 
contain any parametrization of the nuclear equation of state
in terms of an explicit relation between number density, temperature
and the energy density. In equilibrium and in the thermodynamic limit 
($n \to \infty$), however, such a functional relation can be deduced
from the nucleon-nucleon potentials and the cross-sections 
employed in the model.

For the description of the energy per nucleon as a function of density
(assuming $T=0$) usually 
Skyrme type parametrizations (see eq. (\ref{skyrmeansatz})) are used.
This ansatz is phenomenological and can be derived for the
case $\gamma=2$ from the assumption that the
particles interact with each other with two- and threebody contact forces.
It is generalized to effective higher order contact terms by setting
$\gamma>1$ to be a real number.
This generalized ansatz uses three parameters $\alpha, \beta, \gamma$; 
two of them are fixed  
by the constraint that the total energy should have a minimum at the
saturation density
$\rho=\rho_0$ with a value of $E/A=-16$MeV which corresponds to the
the volume energy in the Bethe-Weizs\"acker mass formula.
Together with the condition that a free particle has no binding energy (which
is automatically fulfilled within this ansatz)
there remains one degree of freedom. 
The third parameter is fixed by the nuclear 
compressibility, which is the second derivative of the
energy at the minimum with respect to the density: 
\begin{equation}
\kappa = 9 \rho^2 \frac{\partial^2}{\partial \rho^2} 
\left( \frac{E}{A}\right) 
\end{equation}
  
Two different equations of state are commonly used: A 
hard equation of state (H) with a compressibility of 
$\kappa=$380 MeV and a 
soft equation of state (S) with a compressibility of 
$\kappa=$200 MeV \cite{kru85a,moli85b}. 

%%%%EOS 
To derive an equation of state from the interactions used in
eq. (\ref{vijdef}) we have to convolute the potentials with the
distribution functions assuming an infinite homogeneous distribution. 
In this limit the $V_{\rm Skyrme}$ and $V_{\rm Yuk}$ become functions 
of the constant density only. 
The interaction density of eq. (\ref{rhoint}) as used in eq. (\ref{eosinf}) 
can be replaced by the position independent nuclear matter density.
The integration over the relative momenta of infinite nuclear matter 
Fermi distributions finally turns into a density dependence of the 
momentum dependent interaction. This allows us to obtain  
the compressional part of the nuclear equation of state, which  
depends on the density only. The parameters of the interactions  in 
eq. (\ref{vijdef}) can therefore be chosen that way that a hard or soft eos
is obtained for the infinite matter case. It should again be noted that the 
parameters of the potentials allow a relation to the nuclear equation of state 
(eos) but that the microscopic description works as well for systems far off from
equilibrium where no eos can be defined.

%It should furthermore be noted that the Coulomb forces of the interactions
%would yield a diverging contribution to the total energy of infinite 
%nuclear matter as has therefore not been neglected in the previous discussion.

The interaction range parameter $L$ influences the 
interaction density  (eq. \ref{rhoint}) for finite systems.
For (homogeneous) infinite nuclear matter the density 
(and thus the potential energy) do not depend anymore on the extension of the
gaussian wavepackets.
Thus, the equation of state of infinite nuclear matter is independent of
$L$. In finite matter $E/A$ also depends on $L$. 
Thus even two parametrizations 
which yield the same eos may produce different results for the reaction 
of two heavy ions. Therefore  we have to adjust $L$ to have reasonable
surface properties. 
In order to allow a physical interpretation $L$ should be in 
the order of the size one expects for the range of the nuclear interaction. 
There exists a range of values for $L$, which allows to fix these properties.
Larger values of $L$ increase the effective range of the interaction and thus
lead to some smearing of fluctuations, which are stronger for more located 
wavepackets (small values of $L$).

\JKalt{It should be noted that also in VUU/BUU type models interaction ranges 
are defined in form of the size of a coordinate sphere and the grid size
respectively. However, in these models these parameters are regarded as
technical parameters which should approach zero for the ideal case.
Furthermore these parameters do not explicitly enter into the forces
as it happens for QMD. }

Hence, the nuclear equation of state can only be defined as
the bulk properties in the limit of an infinite system:
The concept of the
nuclear equation of state 
as discussed here does only make sense 
for large macroscopic systems in (at 
least local) equilibrium, while the ansatz with mutual interactions 
has no restrictions with respect to the size
and is therefore also applicable for
finite systems far off equilibrium.
The time evolution of the non-thermal system of two reacting heavy ions 
is completely determined by the two-body potentials and the scattering
cross sections, respectively.

In QMD the parameters $t_1 ... t_5$ are uniquely related to
the corresponding values of $\alpha, \beta, \gamma, \delta$ and $\epsilon$ 
which serve as input. The standard values of these parameters can be found in
table \ref{eostab}.

%%%%%%%%%%%%%%%%%%%%%%%%%%%%%%%%%%%%%%%%%%%%%%%%%%%%%%%%%%%%%%%%%%%%%%%%%%
\subsubsection{Inclusion of collisions}
As stated above the imaginary part of the G-matrix acts like a collision 
term. In the QMD simulation we restrict ourselves to binary collisions 
(two-body level). The collisions are performed in a point-particle sense in a 
similar way as in VUU or cascade:
Two particles collide if their minimum distance $d$, 
i.e.\ the minimum relative 
distance of the centroids of the Gaussians during their motion, 
in their CM frame
fulfills the requirement: 
\begin{equation}
 d \le d_0 = \sqrt{ \frac { \sigma_{\rm tot} } {\pi}  }  , \qquad
 \sigma_{\rm tot} = \sigma(\sqrt{s},\hbox{ type} ).
\end{equation}
where the cross section is assumed to be the free cross section of the
regarded collision type ($N-N$, $N-\Delta$, \ldots).

Beside the parameters describing the $N$--$N$ potential, the 
cross sections constitute another major part of the model. In principle,
both sections of parameters are connected and can be deduced from 
Br\"uckner theory. QMD-calculations using consistently derived cross-sections
and potentials from the local phase space distributions have been discussed
e.g.\ in \cite{jutta}. Such simulations are time-consuming since the 
cross-sections and potentials do explicitly depend on the local phase space
population. 

Within the framework of using free cross section one may parametrize
the cross section of the processes to fit to the experimental data if 
available. 
For unknown cross sections isospin symmetry and time reversibility is assumed.

Alternatively, cross-sections may be obtained from theoretical considerations.
For one particular QMD-version the one boson exchange model has been
employed for this purpose. This has the advantage to have a first handle for
the description of cross sections in the nuclear medium.

If two particles scatter, the direction of the final momenta  will be 
distributed
randomly in such a way that the distribution of many identical collisions 
corresponds to the measured cross section. For elastic scattering the 
distribution is taken from   \cite{cug81}:
\begin{equation}
\frac{\mbox{d} \sigma_{\mbox{el}}}{\mbox{d}\Omega} \sim 
\exp(A(s) \cdot t)\quad,
\end{equation}
where $t$ is $-q^2$, the squared momentum transfer (which also includes the
information on the polar angle) and 
$\sqrt{s}$ is the c.m. energy in GeV.
% and A is given in (GeV/c)$^{-2}$.

\message{kodama-part}
It should be noted that the presented treatment of the collisions may cause 
problems with causality since the particles can interact immediately at a
distance. The collision information is given to both particles at the
same time when they are at closest position. It should also be noted that
the time order of the collisions is determined in a common system of all
particles.  The evolution of the system is propagated with one common clock. 
As it has been already pointed out by Kodama et al. \cite{kodama} the time 
ordering is not unique. Thus the choice of the common referential system
may influence the observables. Normally a system is chosen where the
relative velocities with respect to that system are as small as possible.
Thus BQMD used the nucleus-nucleus CM system as referential system and VUU
and IQMD use the nucleon-nucleon CM system. The choice of the Lab system
as referential system would e.g. cause  for the system Au(1AGeV)+Au b=3fm,
hard eos, an enhancement of the flow (in IQMD $p_x^{dir}$ rises from 
$98 \pm 3$ MeV/c  to $110 \pm 3$ MeV/c in the Lab system ) 
and a reduction of the pion number
(in IQMD $N_\pi$ falls from $64\pm 1$ to $60 \pm 1$ in the Lab system).  

Also the choice of the minimum distance point as collision point can be
motivated within this respect. An earlier collision (e.g. at the point 
when the distance is sufficient to fulfill the distance condition) 
could cause stronger acausalities. It will also reduce the mean free
path and thus enlarge stopping and flow \cite{hart}.

%%%%%%%%%%%%%%%%%%%%%%%%%%%%%%%%%%%%%%%%%%%%%%%%%%%%%%%%%%%%%%%%%%%%%%%%%%
\subsubsection{Pauli blocking due to Fermi statistics}
The  cross section is reduced to an effective cross section by the
Pauli-blocking. For each collision 
the phase space densities in the final states are checked in order to assure
that the final distribution in phase space is in agreement with the Pauli 
principle ($f\le 1$).
Phase space in QMD is not discretized into elementary cells
as in one-body models like VUU, in order 
to obtain smooth distribution functions the following procedure is applied:
The phase space density $f_i'$ at the final states $1'$ and $2'$  is
measured and interpreted
as a blocking probability. Thus, the collision is only allowed with a 
probability of $(1-f_1')(1-f_2')$. If the collision is not allowed the 
particles remain at their original momenta.

The Pauli blockers of VUU and QMD show efficiencies of about 94-96 \%, 
i.e. a single ground state nucleus with Fermi momentum would show a
blocking rate of this amount. In order to reduce the noise of spurious
collisions in ground state nuclei additional conditions allow a nucleon
only to collide with a nucleon of the other nucleus or with a nucleon
that has already undergone a collision. Nevertheless the problem of
Pauli blocking causes a limitation of the calculated system to have not
less incident energy than about the Fermi energy. 
 \message{Pauli-part}

%%%%%%%%%%%%%%%%%%%%%%%%%%%%%%%%%%%%%%%%%%%%%%%%%%%%%%%%%%%%%%%%%%%%%%%%%%
\subsection{Numerical structure}

The QMD model consists of three major parts, namely 
i) the {\bf initialisation} of projectile and target, 
ii) the {\bf propagation} of nucleons, resonances and newly produced 
particles due to their mutual potential interactions, and 
iii) the {\bf hard collisions} according to the energy dependent cross section
for the various channels together with the {\bf Pauli-blocking}.

For the propagation the description of the {\bf potential} (or
to be more exact of the real part of the Br\"uckner G-matrix) 
is of crucial importance.

The solution of the transport equations for the $N$-body distribution function
is done in the following way:
\begin{enumerate}
\item
Projectile and target are initialized. For each of these nuclei the nucleons
initialized according to a distribution $f(r,p,t=0)$. This distribution is
essentially constrained by the requirement to reproduce the ground state 
properties of the two nuclei, i.e.\ radii, binding energies. 
\item
The particles are propagated using Hamilton's equations of motion 
(\ref{hamiltoneq}) with a given Hamiltonian $\langle H\rangle$.
\item
Two particles close in coordinate space may perform a collision. The particles
change their momenta respecting the Pauli principle.
\end{enumerate}

The input into the program may be subdivided into three classes of parameters
\begin{description}
\item[Reaction parameters:] projectile and target masses (and charges), 
bombarding energy, impact parameter.
They define the whole kinematics of a single
event. 

\item[Physics Parameters:] interaction range, potential parameters, in medium 
cross sections and decay widths, etc. They correspond to a detailed description of 
interactions and may be changed within a reasonable range. Finally their
deduction is a particular
goal of the comparison between calculation and experiment.

\item[Technical parameters:] time step size, initial distance, cutoff 
parameters, maximum collision distance, etc. They are used to perform
effective calculations on a computer. The observables should not depend 
on them.
\end{description}

If all these parameters are fixed
the calculation of a single event can be performed in the following 
way:
\begin{itemize}
\item
initialize projectile and target nuclei in their ``ground states'' as 
mentioned above,
\item
propagate the constituents of the system according to their mutual 
potential and hard scattering interactions, this includes
\begin{itemize}
\item
calculation of interaction  densities, forces and the Hamiltonian
\item
propagation of all particles according to Hamilton's equation of motion
\item
perform all collisions within this time step. Decide for each collision 
whether its final state is Pauli-blocked. If this is the case: 
keep momenta of collision partners unchanged, otherwise change momenta
according to the angular distribution of this particular channel.
\end{itemize}
\item
output of information (coordinates, momenta, scattering partners, \ldots)
about the intermediate reaction stages and output of the final 
 phase-space configuration (which would correspond to the freeze-out 
configuration in a thermal picture).
\end{itemize}
This procedure is repeated until sufficient statistics,
i.e.\ a large number of independent events, is obtained.

This principal structure is common for all QMD realizations,
which differ, however, in details and the initialisation of projectile
and target. In the following we will study the influence of these
differences on observable and nonobservable quantities.

%%%%%%%%%%%%%%%%%%%%%%%%%%%%%%%%%%%%%%%%%%%%%%%%%%%%%%%%%%%%%%%%%%%%%%%%%%
\section{Description of particular QMD model realizations}
The original QMD \cite{ai85a,ai86} program was developed further 
to include momentum dependent interactions \cite{ai87b,pei89}.

\subsection{BQMD}
The original QMD has been rewritten
by Bohnet et al.\cite{boh91} for the purpose of studying low energy 
fragmentation data. This program has been dubbed BQMD 
since it was designed for describing the proper binding of a nucleus 
in order to describe 
fragmentation processes\cite{beg93,muell93,jeong,goss94}. 
An improvement on the stability against artificial particle evaporation 
has been achieved in BQMD by a procedure explained below which causes
fluctuations of the energy around the mean value by 2 MeV/nucleon.  
\JKalt{
Proper binding energies have been achieved in BQMD on the cost of
large dynamical fluctuations in the initial state and by a moderate energy
non-conservation ($\approx$ 2 MeV/nucleon for single nuclei)
on an event by event basis.
Its particularities are due to the initialisation and the potentials,
which will be explained in more detail below.
}

\subsubsection{Initialisation in BQMD}
In BQMD the nucleons are distributed within a sphere 
with a Wood-Saxon-type density profile. (The original QMD used a 
sphere for the distribution of the centroids of the Gaussians.) 
The maximum Fermi-momentum is limited by the local
binding energy of the nucleon in order to keep all particles bound. 
By this procedure, however, the mean kinetic energy of the particles is lowered
to about 10 to 12 MeV/nucleon.
The ground state central density is assumed to be $\rho_0=0.15 fm^{-3}$. The Gaussian
width for the interactions are chosen to be $L=4.33 fm^2$.
The binding energy as given in the Weizsaecker mass formula 
is reproduced from Lithium up to the heaviest nuclei 
\cite{ai91}. 
%The proper binding energy garanties a very good stability of the
%ground state nuclei.
As already seen in fig. 10 of \cite{ai91} this particular version suffers
from fluctuations of the rms radius. The consequences will be discussed
later.

\subsubsection{Potentials in BQMD}
The range and the strength of the Yukawa potential in BQMD has been 
chosen to describe the surface of the nucleus best. 
In order to keep the nuclear equation of state and the 
binding energy independent of the
Yukawa interactions and to keep 
the binding energy at its experimental value, the 
coupling constant $t_1$ of the Skyrme-type two body interaction.
is modified according to \cite{ai91}
\begin{equation} \label{yukred}
 t_1^i (i) \rho(\vec{r_{i0}}) =
 t_1 \rho(\vec{r_{i0}}) -
\sum_{j} U_{ij}^{Yuk} \,.
\end{equation}
Note that the Skyrme and Yukawa
coupling constants are different for each particle here.
With this procedure the validity of Newtons theorem {\em actio = reactio} 
can be assured on the ensemble average only, which also leads to violation
of energy conservation in single events.
The energy fluctuates about 2 MeV/nucleon around
the mean value \cite{ai91}.  
The range of the Yukawa-potential is chosen as 1.5 fm.

\subsubsection{Collision term in BQMD}
BQMD has in common with the 
original QMD that it uses nucleons and deltas only. 
%Isospin degrees of freedom are now regarded concerning the
%collision term. The cross sections for proton/neutron and proton/proton 
%are distinguished (the neutron/neutron cross section is assumed to be like 
%proton/proton).  Protons and neutrons are treated equivalent concerning 
%the potentials. 
%Conce
The employed cross sections have been parametrized by Cugnon 
\cite{cug81}.
All nucleons interact with the same average cross section without distinction 
in
isospin. The elastic cross section is given by a constant value of 55 mb for
collisions with $\sqrt{s} \le 1.8993$ GeV and for higher energies by the 
parametrization:
\begin{equation}
\sigma_{el}(mb) = \frac{35}{1 + 100 \cdot(\sqrt{s}/GeV -1.8993)}+20
\end{equation}
The inelastic cross section $NN \to N\Delta$ is zero for $\sqrt{s}\le 
2.015$ GeV
and for higher values by the parametrisation
\begin{equation}
\sigma_{in} (mb) = \frac{20 x^2}{0.15-x^2} \qquad x =\sqrt{s}/GeV -2.015
\end{equation}

The angular distribution of the collisions is described by
$\frac{\mbox{d} \sigma_{\mbox{el}}}{\mbox{d}\Omega} \sim \exp(A(s)\;t)$
with

\begin{equation}
A(s)\,=\, 6\, \frac{(3.65\; 
(\sqrt{s}/GeV - 1.8766))^6}{1 + (3.65 \;(\sqrt{s}/GeV - 1.8766))^6}
\quad .  
\end{equation}

\subsection{IQMD}
The Isospin-QMD (IQMD) \cite{ha89,hart} treats
the different charge states of nucleons, deltas and pions 
explicitly, as inherited from the VUU model.
IQMD has been used for the analysis of collective flow effects of 
nucleons \cite{ha89,mpl,chplb94,ssoff} and pions \cite{ha88,baprl,baprc}.
Comparisons to experimental data with this model have been presented in
\cite{madey,claesson,ramilien}. 
As it has been developed from the VUU-model, its coding is therefore
independent of the
original QMD. The isospin degrees of freedom enter into the cross sections 
(here cross sections of VUU \cite{kru85a} similar to the parametrizations
of VerWest and Arndt \cite{vw82} have been taken, see also
ref. \cite{baprc}) as well as in the Coulomb
interactions. The elastic and inelastic cross sections for proton-proton
and proton-neutron collisions used in IQMD are shown in figure \ref{hipsigma}.
The cross section for neutron-neutron collisions are assumed to be equal to the
proton-proton cross sections.

\subsubsection{Potentials used in IQMD}
The IQMD-model offers rather stable density distributions and good
energy conservation, however for the price of nucleon evaporation and
and improper binding energies ($E_{bind}\approx 4-5$ MeV/nucleon for heavy
nuclei instead of 8 MeV/nucleon).

 %(eq. \ref{H-qmd})
In addition to the use of the explicit charge states of all baryons and mesons
a symmetry potential between protons and neutrons 
corresponding  to the Bethe-Weizs\"acker mass formula has been included 
\begin{equation}
V^{ij}_{sym}= t_6 \frac{1}{\varrho_0}
 T_{3i} T_{3j} \delta(\vec{r}_i - \vec{r}_j) \quad t_6 = 100\, \rm MeV
\end{equation}
where $T_{3i}$ and $T_{3j}$ denote the isospin projections of particles
$i$ and $j$. 
Other baryonic potentials like $V^{ij}_{\rm Skyrme}$ and 
$V^{ij}_{\rm mdi}$  are defined isospin-independent like in all other 
flavors. The Yukawa potential in IQMD  $V^{ij}_{\rm Yuk}$
is very short ranged ($\mu=0.4$fm in contrast to $\mu=1.5 fm$ in BQMD) 
and weak. The modification of the $ \alpha $ term of the static potential
is done in an particle independent
way. As in BQMD this corresponds to the interpretation that an 
additional term in the 
Skyrme ansatz which is proportional to $(\nabla \rho)^2$ can be expanded 
in first order to a term linear in density (which reduces $\alpha$ 
effectively) plus Yukawa potentials. Additional attractive Yukawa forces hence
modify the EOS (and therefore the $\alpha$ term has to be modified
to obtain the same EOS). Yukawa forces stabilize the nuclei because 
of the increase of the interaction range as compared to a 
$\delta$-like Skyrme-potentials.
Thus nucleons notice earlier that they will arrive at the surface and are
more effectively decelerated as without this
potential. In addition the fluctuations are reduced.
  
\subsubsection{Pions in IQMD}
Free pions are moving under the influence of the Coulomb interactions.
Pions may be produced by the decay of a 
$\Delta$-resonance and may be reabsorbed by a nucleon
forming a delta again. 
IQMD and HQMD, which will be described in the next section,
differ concerning the pion production in the production cross
sections (HQMD uses cross sections based on the one boson 
exchange model), the included resonances
(HQMD contains additionally $N^*$ and $NN \rightarrow \Delta \Delta$
collisions) and the angular 
distribution of inelastic
collisions (HQMD has more 
realistic non-isotropic distributions obtained from OBE calculations
which are not present in original IQMD).
Recent updates of IQMD calculating the pion production 
(e.g. \cite{baprl,baprc}) also use the 
inelastic angular distributions of HQMD. The effect of this modification
on nucleonic observables is quite small.

\subsubsection{Initialisation in IQMD}
The most important difference to BQMD is the initialisation. 
In IQMD the centroids of the
Gaussians in a nucleus are randomly distributed in a phase space sphere
($r \le R$ and $p \le p_F$) with $R=A^{1/3}\cdot 1.12$ fm corresponding
to a ground state density of $\rho_0 =0.17 fm^{-3}$. The Fermi momentum
$p_F$ depends on the ground state density. For $\rho_0=0.17 fm^{-3}$ it has
a value of about $p_F \approx 268$ MeV/c. 
While, as said, in BQMD the maximum momentum is determined by the local
binding energy (which causes an effective reduction of the total Fermi energy
to about 10 -- 12 MeV), in IQMD the momenta are uniformly distributed within a
momentum sphere $p \le p_{Fermi} \approx 268 MeV/c$ without further local
constraints. Therefore it may happen that nucleons close to the surface, 
where the local potential energy is low, are unbound initially. 
This possibility is not given in BQMD or HQMD.
It gives, however, a reduced binding energy per nucleon
as compared to the Weizs\"acker mass formula.
Hence the initialized nuclei are less stable against spurious particle
evaporation as compared to BQMD.
On the other hand this ansatz makes available the full Fermi-energy
calculated from the Skyrme ansatz. The full Fermi pressure yields (as
compared to BQMD)  
a stronger stability of the density profile  against vibration modes. 
Finally it should be noted that
IQMD performs a Lorentz contraction of the nucleus coordinate distribution 
which is not present in BQMD and which becomes important for higher energies
$E/nucleon > 1$ GeV. %% \cite{}.

\subsubsection{Interaction range}
As it has already been stated, the Gaussian width can be regarded as a
description of the interaction range of a particle. Its influence
disappears for infinite nuclear matter whereas for finite systems it may
play a non negligible role.  

In IQMD the Gaussian width can be used as an optional input parameter.
The default version of uses a system dependent Gaussian width while BQMD uses
$L=4.33\,fm^2$ independent of the system size. The system dependence of $L$ in
IQMD has been introduced in order to obtain maximum stability of the nucleonic
density profiles. 
As an example for Au+Au  a value of $L= 8.66\,fm^2$ is choosen, for Ca+Ca 
and lighter nuclei $L=4.33$.

\subsection{HQMD}
HQMD is an upgrade of QMD which combines optional features of 
BQMD and IQMD. 
It does not remedy the shortcomings of BQMD and IQMD, but allows to
study the influence of the different modules on physics results.
In addition, higher resonances ( the $N^*$(1440)), 
free pions and the proper isospin coupling have been incorporated by 
Huber et al. \cite{hub94}. The isospin degrees of
freedom play an important role especially for the particle
production. The employed inelastic cross sections 
$ NN \rightarrow NN^*, N\Delta $ and $ \Delta \Delta$
have been calculated within an one-boson
exchange model (OBE). Also the angular distribution of the inelastic reactions
was calculated and parametrized in the following way:
\begin{equation}
\frac{\mbox{d} \sigma_{\mbox{in}}}{\mbox{d}\Omega} \sim a(s) \; 
			\exp(b(s) \cdot \cos \theta)\quad,
\end{equation}
$a(s)$ and $b(s)$ are functions of $\sqrt{s}$ and vary in their 
definition for different intervals of $\sqrt{s}$ (see table \ref{inelast}).
$\theta$ is the polar angle.
It should be noted that VUU, BUU and IQMD (in its older version) assumed 
isotropic scattering for the inelastic channels which causes differences in the
flow at higher energies.
For elastic collisions a new parametrisation \cite{cug89} has been used which
can be taken from table \ref{sigcugnew}.

In addition the numerical propagation routines have been changed to a 
higher accuracy. A 4th order Runge-Kutta propagation scheme allows an energy 
conservation of about 1 per mille.  The Yukawa interaction has been
suppressed.

This upgrade of QMD which has been dedicated to the 
question concerning the meson production. It is quoted as HQMD 
because it contains  higher resonances.

HQMD offers the possibility to choose between the two initialisation
modes of BQMD and IQMD. Moreover one can choose between the different 
parametrizations of the cross section as described above: the cross section 
parametrization used in IQMD and that used in BQMD. It was checked that
it reproduces the results obtained with BQMD and with IQMD
if the corresponding subroutines are used. Therefore it may be 
directly used to analyse the effects of the different ingredients in the 
QMD flavours.

\subsection{Other flavours}
There exist several other flavours of QMD. 
Peilert et al. use an extension of QMD with additional implementation of a 
so-called Pauli
potential \cite{pqmd}. These models use a strong repulsive potential which 
depends on the distance of the particles in phase space. 
It is effective in momentum- and in configuration
space and prevents two identical particles from coming too close in
phase-space. Its parameters have been adjusted to the temperature- and
density-dependence of the energy per particle of an ideal Fermi-gas
\cite{Dorso87,Peilert91}. 
With aid of such a potential selfconsistent nuclear ground
states for nuclei with $N$ neutrons and $Z$ protons as well as for
infinite nuclear matter
can be constructed by searching for the minimum in the multi-dimensional
potential-energy surface of $N$ neutrons and $Z$ protons. The nucleons carry
their proper Fermi-momentum, however due to the momentum-dependence of 
the Pauli-potential, their velocities (=kinetic momenta $\partial H/\partial p$)
vanish in the ground state. This can be interpreted as a first approximation 
to antisymmetrization in finite nuclei on the two-body level. 
However one should note that for the scattering of individual nucleons a
Pauli potential and antisymmetrisation yield different effects.
Konopka et al. generalized
this concept by treating collisions of Gaussians instead of point particles as
it is the case in all other QMD versions. For the sake of numerical feasibility 
the cross section has to be assumed as constant. 
This model has been utilized
for analyses of the FOPI-data at lower energies \cite{konopka}.
The basic differences in the observables calculated with the 
Pauli-QMD and with IQMD can be explained by the use of a isotropic
41 mb cross section in the Pauli-potential QMD. However, one should
stress that the Pauli-potential has - due
to the strong momentum dependent potential -  a different physical input
as compared to all flavours discussed and hence a detailed comparison
is not intended in this paper. 

Further studies with QMD have been done by Jaenicke et al. who replaced
potentials and cross sections of BQMD by those calculated from 
a Br\"ucker G-matrix \cite{jutta}. Comparison of experimental data with 
this model has been performed by the FOPI-collaboration \cite{ramilien}.
 
Lehmann and Puri extended HQMD by including a relativistic
covariant propagation scheme of the RQMD-type. 
The physical inspiration of the scheme was taken from 
the RQMD -model of Sorge \cite{Sorge}, which originally was footed on the
IQMD and vastly extended for the description of high energy collisions in the
potential and the collision parts. 
Similarly, the RQMD of Lehmann and Puri is an 
numerical extension of the HQMD model. 
The inclusion of a
covariant treatment of initialisation, Pauli-blocking and potentials 
yield at high energies ($E/A > 1$GeV) 
some differences to the `nonrelativistic' HQMD
which are described in \cite{leh95,pu95}. 
The relativistic and the nonrelativistic version 
agree at low beam energies. Hence this program
allows a systematic investigation from very low energies to 
very high energies. The required computing time, however, is one order of
magnitude higher.

Kaon production has been intensively studied \cite{aijae} using a modified version of
HQMD. The differences are the neglect of free pions 
and the parametrization of the inelastic cross sections where only
the reaction channel $NN \rightarrow N\Delta$ has been employed.  
Differences in the kaon production between this version and IQMD 
have been discussed in \cite{kaon94}.

%%%%%%%%%%%%%%%%%%%%%%%%%%%%%%%%%%%%%%%%%%%%%%%%%%%%%%%%%%%%%%%%%%%%%%%%%%
\section{Numerical test and results}
The above discussed QMD versions allow for simulations of heavy ion
collisions up to $\approx 2$ GeV/nucleon. Above this energy higher resonances, which are not included
in the models under consideration become more and more important.
The model gives detailed information about all one-body observables, such as 
single particle spectra, and many-body observables, such as particle
correlations and fragment formation, on an event-by-event basis.
Thus the structure of these theoretical data is analogous to
experimental data. 
The independent development of BQMD and IQMD 
including different model assumptions lead to different results in some cases. 
In this section, we therefore
compare several QMD versions with particular attention to some standard
observables. It is demonstrated that most of the differences are
related to the different treatment of the initialisation of the two
colliding nuclei.

\subsubsection{Rapidity distributions}
A quantity, which is crucially related to the possible 
formation of a thermally equilibrated source is the
rapidity distribution of baryons. 

Fig. \ref{vuu0dndy} shows the rapidity distribution $dN/dy$ of nucleons  
in the reaction Au+Au, b=3fm at 1 GeV/nucleon incident energy.  
IQMD (squares, dotted line), BQMD (circles, full line) 
and VUU calculations (triangles, dashed line) 
using a hard equation of state without momentum dependent
interaction give quite similar results.
BQMD shows a slightly broader distribution than IQMD and VUU.
As it has been already stated in ref. \cite{ai87b} the rapidity distribution 
depends strongly on the collision term and only slightly on the used 
nucleonic potentials. From this we can 
conclude that the hard collisions do not lead to large differences.

The remaining differences of about 40 units at midrapidity divides
up as follows (the statistical error of each of the midrapidity dN/dy values
amount to about 10-15 units):
The change from large to small width enhances the value by about
20 units, whereas the different cross sections contribute a lowering by 30
units when switching from IQMD to BQMD. The BQMD initialisation lowers
by about 20 units and the Yukawa potential in BQMD reduces by about 10
units. 

At lower bombarding
energies the dynamics is no longer dominated by the hard collisions
and the nucleon potential becomes more important.
At 150 MeV/nucleon, the rapidity distributions exhibit some larger differences
between the two QMD version used. This, however, is due to the
inclusion of Yukawa forces in BQMD. The results look more alike if
this term is omitted in both calculations.
The dN/dy at midrapidity reaches about 11\% smaller values in BQMD, which are
decomposed as follows: 
7\% enhancement due to the interaction range, 
9\% reduction due to the cross sections,
5\% reduction due to Yukawa, 
and a slight reduction ($\approx$ 2\%) due to different initialisations.

%Other variables like stopping, pion and 
%kaon number show a much weaker dependence on the density profile fluctuations.
%We take advantage of HQMD which allows this comparison
%and display in fig. \ref{hip7dyp} 
% the  calculated in HQMD with
%BQMD-initialisation and IQMD-initialisation. With the former initialisation
%we observe but the rapidity distribution is about the same. The increase
%of the pion number is due to the larger mean free path for the process 

\subsubsection{Transverse flow}
Let us now focus on an observable whose investigation is  
strongly motivated by its
dependence on the nuclear equation of state \cite{st86,moli85b,pei89}
(besides its dependence on the collision term and on the centrality), 
namely the transverse flow in plane. This variable
turns out to be extremely sensitive to a lot of parameters as we will see.
The amount of transverse flow created in heavy ion reactions is known
as a measure of the pressure built up during the reaction and it thus
can provide information about the underlying equation of state.%% \cite{}. 

Fig. \ref{vuu0pxe} compares the excitation functions of flow
for VUU, IQMD and BQMD with their default width parameters 
$L=8.66\,fm^2$ and $L=4.33\,fm^2$ respectively. 
It is found that  VUU and IQMD show a similar
behaviour with a rise of the flow up to 1 GeV incident energy 
(which is also in good agreement with experimental data) while
BQMD shows rather weak rise of the flow. This weak rise is in 
disagreement with experimental flow data.
The reasons of the differences between BQMD and IQMD  
shall be briefly investigated. 
\JKalt{
It should, however, be kept in mind that
the flow results of BQMD should be taken with caution due to the instabilities
of the flow variable.}

The transverse flow is not only sensitive to the repulsion of the
compression zone formed 
by excited nuclear matter, but also to surface properties, such as
the range of the nuclear interaction. 
This quantity may be varied
within the QMD approach in two different ways: 

The range of the 
optional Yukawa force in QMD is an adjustable parameter, it can be 
used to stabilize the width of the nuclear surface of a given density
profile. The width parameter $L$ of the gaussians
serves as an effective interaction range as well.
It should be noted that a change of the interaction range also changes the
density gradient in inhomogeneous systems (this can be demonstrated
by regarding the density profile of a 'box') and therefore directly enters
into the gradient of the potential.
It was found that default  BQMD calculations with a Yukawa potential
yielded a  directed transverse momentum,
$p_{x}^{dir}$ which is about 10 MeV/c higher than for calculations 
where only the Skyrme interaction is used. This is due to the 
fact that a finite range Yukawa smears out the potential gradient more than 
a $\delta$-function and hence reduces the force in transverse direction.

In IQMD the inclusion of Yukawa forces does not give significant effects 
on the nucleonic flow. It should however be noted, that in IQMD the range 
of the Yukawa force is smaller (0.4 fm as compared to 1.5 fm of BQMD) and
that actio=reactio is respected for the two-body interactions.

Both models agree in the observation that a
broadening of the Gaussian width  $L$ reduces the 
flow. This also corresponds to the fact that the density gradient
to the high density region is smeared out.  

The influence of the interaction range on the flow can be studied in
fig. \ref{hipl4px} which compares IQMD results of the flow for 
$L=8.66\,fm^2$ and $L=4.33\,fm^2$.
In IQMD the default value for Au+Au is chosen to be $L=8.66\,fm^2$.
A smaller interaction range enhances the flow value by about 10 MeV/c
at 400 AMeV and by 20 MeV/c at 1 AGeV.
A further difference caused by the interaction range is the
density of the saturation of the potential, i.e.the density where the
potential supports maximum stability of the initial state versus 
vibration modes. For the IQMD initialisation the maximum stability
is reached for $L=8.66 \,\rm fm^2$ at about $\rho=0.17 \rm fm^{-3}$
and at about $\rho=0.15 \rm fm^{-3}$ for $L=4.33\, \rm fm^2$.

The differences in the flow results between BQMD and IQMD motivated to
search for parameters which might influence the flow.
There are three major differences between BQMD and IQMD calculations:
Besides the interaction range they concern 
the initialisation, the cross-sections employed and the
different values for the saturation density. For a better comparison 
we changed in the following the interaction range of IQMD to $L=4.33\,fm^2$.
As a first step we investigate the dependence on the initialisation 
of the nuclei, which also includes the role of the value of $\rho_0$.
The BQMD and IQMD initialisations differ in three aspects:
a) the shape of the coordinate distribution of the particles,  
b) the average central density in the nucleus and 
c) the limitation of the Fermi momentum to the value obtained by a 
local density approximation or the full Fermi momentum, respectively.

The dependence on each of these differences is studied in fig. 
\ref{hip11pxe} which shows results of a HQMD calculation with
inclusion of modules from BQMD and IQMD. It demonstrates that the very 
same dynamics, i.e.\ same 
forces, same cross-sections and same equations of motion lead to 
considerably varying results depending on the initial conditions chosen.

Here IQMD-ini (squares) denotes the default IQMD initialisation with
$L=4.33 fm^2$. 
using hard sphere for the centroids, no constraint to the Fermi-momentum 
and $\rho_{ini}=\rho_0=0.17 fm^{-3}$.
The diamonds describe a calculation with a different initial  density
$\rho_{ini}=0.15fm^{-3}$. We see a reduction of the flow at highest densities.
This effect is known from hydrodynamical studies using the
Rankine-Hugoniot equations.\cite{hstprivat}.
It should be noted that the values obtained with $L=4.33$ obtained at
$\rho_{ini}=0.15fm^{-3}$ (i.e. at maximum stability) are comparable 
with the values obtained with $L=8.66\, \rm fm^2$ and $\rho_{ini}=0.17 fm^{-3}$
i.e. at maximum stability for the $L=8.66 fm^2$ case.

The triangles denote a calculation where additionally the Fermi momenta are
constrained by the binding energy (similar as done in BQMD). Here we
see an enhancement of the flow at low energies. Besides the flat shape
of the excitation function this constraint on the Fermi momentum also
causes strong fluctuations of the 
rms radii \cite{ai91}. As a consequence - as we will see later - this 
introduces a considerable systematic error of the flow values.
%of the flow value with respect to the 
%initialisation distance.

BQMD-ini (circles) finally denotes the BQMD default 
with a Wood Saxon distribution, 
the local constraints of 
the Fermi momentum and the saturation density of 
$\rho_{ini}=0.15 fm^{-3}$. Since the density profile is now smeared out even more
an additional reduction of the flow can be found at high energies.

The composition of all three effects causes the BQMD-initialisation
to yield a very flat excitation curve while for the IQMD-initialisation a
strong dependence of the flow on the incident energy is observed.  

It should be noted that similar effects as reported for the flow 
are also found for the particle production. These effects are weaker but
lead to the same picture. Effects that simulate a weaker repulsion and
thus cause a weaker flow will yield an enhanced particle production.

\JKalt{
In fig.\ref{hip21pxe} the dependence of the flow on the collision term
is studied. Again we use HQMD but employ collision terms as in the
BQMD and IQMD simulations. We see that the influence of the collision
term on this observable is much smaller than that of the
initialisation, however not negligible. If collisions are performed
like in IQMD we get about 10\% larger $p_x^{dir}$ values as compared
to the cross section employed in BQMD. This is the only variable 
where we have found differences which are exclusively due to the 
differences between the employed cross section.
}
The collision term also influences the excitation function of flow.
It was found that the collision terms of IQMD and HQMD yielded about
the same values while the BQMD collision term causes a decrease of about
5-10 MeV/c relativ to the IQMD collision term. 

The different $p_x^{dir}$ values reflect themselves in different
dependences of $p_x$ on the rapidity. Figure \ref{vuu0pxy} 
compares the transverse momentum in the reaction plane $p_x(y)$
for the system Au(1AGeV)+Au at b=3fm impact parameter. 
We see that IQMD gives values similar to VUU while  
BQMD yields much lower flow values close to beam and target rapidity.
It should furthermore be noted that similar effects have been found in the
analysis of the flow out of plane. BQMD yields a less pronounced squeeze-out 
as compared to VUU and IQMD.

In conclusion it is found that the description of the flow depends 
strongly on the detailed description of the initial state of the nuclei 
as well as on the interaction range. The constraints on the Fermi momentum
as used in BQMD lower the Fermi pressure and yield a 
nearly flat excitation function of the flow which is in
contradiction to current data.

\subsubsection{Fragment production}
Another key issue for heavy ion reactions is the
simultaneous production of several intermediate mass fragments, i.e.\
clusters which are heavier than
$\alpha$'s but considerably lighter than typical fission products.
This phenomenon, usually referred to as multifragmentation, has lead
to numerous speculation that this may be the signal for the occurrence
of a liquid vapor transition in nuclear matter \cite{pochodzalla}. 

A lot of studies involving QMD models addressed this
issue as well. QMD does not explicitly include a phase
transition, even more so, it is a non-equilibrium transport theory,
where equilibrium need not necessarily be established to be applicable
to multifragmentation reactions, as it is the case in statistical 
models for nuclear fragmentation \cite{botvina}.  

It should be noted that there exist problems in describing the properties
of Fermi systems at low temperature \cite{pqmd,donangelo}. However, it should
also be noted that the fragment distributions obtained with QMD are in the
range of the different statistical models \cite{muell93}. The difference
between the distributions from QMD and these models is in the same order
the differences between those models themselves.
\message{rev-frag}

QMD predicted the emission of several fragments in a single event,
qualitatively similar to the experimental observations and at lower
energies also quantitatively \cite{boh91,gossiaux951,gossiaux952,beg93}. 

A realistic description of fragmentation processes within QMD is one
of the most complicated tasks. At higher bombarding energies
($E > 400$ MeV/nucleon) fragment
formation is already a rather rare process. At lower energies, 
where multifragmentation is a major reaction channel (between 50 and
200 MeV/nucleon), the reaction is slowed down, which requires a
improved accuracy of the calculation.

One aspect is that single nuclei at rest also start to evaporate
nucleons and fragments after 50--100 fm/c. This effect has to be
minimized which sets stringent conditions on the stability of single
nuclei. Moreover the nuclear binding needs to be properly described.

At 50 MeV/nucleon beam energy for a symmetric system, each nucleon 
carries 12.5 MeV kinetic energy in the center of mass. Together with
a binding energy of about 8 MeV per nucleon, only 4.5 MeV/nucleon are
available in the center of mass. This has severe implications on the
required accuracy of the description of ground state nuclei. If the 
binding energy is missed by only 1 MeV/nucleon, then a 22\% 
different total energy is used in the calculation. At 100 MeV/nucleon this 
uncertainty still amounts to 6\%.

One crucial aspect, as far as the fragmentation properties of QMD are
concerned, is the interaction range which is directly related to the
width of the gaussian wavepackets. More extended wave packets i.e. \ 
a long interaction range leads to a smaller number of fragments.
These fragments are somewhat heavier than those fragments from
simulations with smaller wavepackets. This behavior has essentially
two reasons: in the case of broader gaussians, particles in a cluster
are bound to a larger number of other nucleons inside the cluster.
On the other hand, with a smaller width the fluctuations are enhanced
and an excited nucleus dissolves more easily.

For example, BQMD with a Gaussian width of $L=4.33$ fm$^2$ gives 12.7
IMFs in Au (150 MeV/nucleon) + Au
at b=3 fm. IQMD with more extended gaussians ($L=8.66\,fm^2$) yields 6.6
IMFs only. It should be noted that in the present analysis the charge 
has not been regarded (especially since BQMD has no explicit charges).
Therefore we used $5\le A \le 19$ for the numbers obtained above. 
However, if we employ the same Gaussian width for both models we
obtain almost the same results. This can be seen in Fig. \ref{hip15a7}
where the fragment mass spectra have been compared for both models
using both interaction ranges.

Parameters other than the interaction range, e.g.\ the value of the
saturation density or the different treatment of Fermi momenta do not
affect the intermediate mass fragment multiplicity significantly.
We also find no differences on the different cross sections or potentials.

The range of Yukawa forces do not significantly
influence the mass distribution of the fragments
as it was found for the Au(150AMeV)+Au at b=3fm. This observable shows
only dependence on the Gaussian width.

In conclusion it can be stated that the interaction range shows strong 
significance on the fragment production.
The smaller values of $L=4.33\,fm^2$ used in BQMD show much better agreement
to existing data as $L=8.66\,fm^2$ used in IQMD. Furthermore BQMD shows  a 
better stability against particle evaporation and better binding energies. 
This is due to the constraints on the Fermi momentum as well as to the
Yukawa potentials. The vibration modes resulting from the Fermi momentum
constraints do not show strong influences on the fragmentation, at least
in central collisions. Although the initialisation does not have a strong 
influence on the fragmentation pattern in central collisions, an initialisation
which combines both, a proper binding energy (as it is achieved in BQMD)
and a proper density profile (as it is done in IQMD) is preferred.
This achievement is one of the main design goals of a new molecular dynamics
scheme of the QMD type \cite{uqmd}.

\subsubsection{Particle production}
Let us now turn to the question of particle production.
In the regarded energy domain mainly the production of 
pions and subthreshold kaons is of interest.

Concerning the description of pion production  
the results of BQMD are not regarded, 
since it has no free pions and the cross section  parametrization was
not adapted to pion physics. Instead we will compare HQMD and IQMD.
It is found that the cross section parametrizations of HQMD and IQMD
yield very similar results. The change of the interaction range changes the
pion multiplicity by only 5--10\%.
A strong influence can, however, be obtained 
from the initialisation procedure.
As an example, rapidity distribution of pions are 
displayed in Fig.\ \ref{hip7dyp}. The two calculations differ only
in the initialisation, forces and cross-sections are identical in
both cases. The BQMD initialisation (that with the lower density) 
yields about the same shape of the pion rapidity distribution as 
a calculation using the IQMD initialisation. The absolute number of
produced pions is about 30-40\% 
larger in the case of the BQMD initialisation.

The difference cannot be explained neither by Pauli blocking nor by 
absorption effects.
It is found that although the calculation with the IQMD-initialisation
yields higher densities, the calculation with a BQMD-initialisation shows
higher collision numbers. This corresponds to the fact that up to maximum
compression calculations with a BQMD-initialisation  loose compressional 
energy (and thus gain kinetic
energy) while calculations with an IQMD-initialisation  
gain compressional energy and loose kinetic energy.

The arguments used in the discussion of the flow (see Figure \ref{hip11pxe})
still hold for the particle production. The different parts 
of the initialisation which cause an increase of the flow yield 
correspondingly a decrease
of the pion number. For Au(1 AGeV) +Au $b=3$fm the change of the 
initialisation density from $0.17$
to $0.15\rm fm^{-3}$ yields (for $L=4.33\rm fm^2$) an increase of
about 20 \% 
and the change from a hard sphere initialisation to a Woods-Saxon type
initialisation another enhancement of about 15 \%. 
The constraints on the Fermi momentum yield no visible  influences on the pion
number at 1 AGeV energies.

It should furthermore be stressed that all regarded models (VUU, IQMD, HQMD)
perform a delta decay using a lifetime which is the inverse of the 
mass-dependent decay-width. Changes concerning this description might
have strong effects on the rapidity distributions. %%% \cite{}.

The study of subthreshold kaon production is motivated by the strong 
dependence of the kaon multiplicity on the nuclear eos
\cite{ai87b,kaon94,aijae,Lang92}. 
It has been found that a hard equation of state yields a stronger repulsion
and lower densities of the compression region than a soft eos \cite{peniscola}.
Therefore a hard eos shows stronger flow and a smaller kaon 
multiplicity. The pion multiplicity shows only slight dependences on the eos
since both equations of state yield about the same compression 
densities \cite{st86,peniscola}.

Kaon production have not been studied within the BQMD model but with an upgrade version
called QMDRKNC \cite{aijae}. This version does not include free pions, therefore
the deltas have an infinite lifetime. A first comparison of this version
with IQMD has been presented in \cite{kaon94}. 
The kaon numbers obtained with both models agree for Au (1 AGeV) +Au within
20-30\%. HQMD with $L=4.33\rm fm^2$ and BQMD initialisation yields
about the same values. HQMD differs mainly from QMDRKNC in the  lifetime 
of the deltas. 
The similar multiplicities of IQMD and QMDRKNC are a result of counterbalancing
 effects which will be briefly discussed:

The  infinite lifetime of the delta in BQMD and QMDRKNC 
causes an enhancement of about 10 - 20\% 
of the kaon number when compared to HQMD. This is due to the dominance of the
channel nucleon+delta $\to$ nucleon+hyperon
+kaon for the subthreshold production. An infinite lifetime enhances the
possibilities for nucleon-delta collisions. Similar numbers have been
found when comparing default IQMD with an calculation with infinite
delta lifetimes.

The BQMD initialisation yields  an enhancement
of the kaon  production by about 20 - 30 \% 
as compared to the IQMD initialisation. 
The reason is presumingly similar to that for the pion production.
The BQMD initialisation allows higher kinetic energies of the nucleons
in the compressed state.

The choice
of a short Gaussian width ($L=4.33\rm fm^2$, BQMD default)
causes a reduction of the kaon number of about 30 \%  as compared to a 
calculation with $L=8.66\rm fm^2$ which is the IQMD default for Au. 
The reason for this may be connected to the argument used for explaining 
the enhancement of flow when using a short width. The density gradient
gets steeper when the interaction range decreases. This simulates a 
stronger repulsion of the compressed nuclear matter.

\subsubsection{Initialisation and Stability}

One of the  seminal problems of the simulation of a heavy ion reaction is the 
proper description of ground state nuclei. One cannot expect that a
reaction is reproduced properly if projectile and target do not have
the observed properties, in particular the proper ground state density.

As we have stressed several times before the choice of the initial 
condition is crucial for a proper description of various phenomena.
Fermi-momenta treated semiclassically as a 
random motion of nucleons inside a nucleus induces significant 
fluctuations of the density profile, if the motion of a single
nucleus is followed for some time.

Fig. \ref{hiptrms4} shows  a time evolution of
the root mean square radii of single Au  nuclei in coordinate and 
momentum space. Concerning the coordinate space
we observe an expansion mode in IQMD and an oscillation mode in BQMD.
IQMD shows best stability if potentials corresponding to a hard eos is
used whereas BQMD shows best stability if a soft eos is used.
The rms radii obtained with BQMD soft eos correspond to the results
presented in \cite{ai91}.
The other eos yield larger fluctuations for both programs.
This stresses once more the fact that a semiclassical approach can be
optimized to a desired purpose but on the cost of other applications.

The fluctuations of the rms radii in momentum space demonstrate that
the potentials are not saturated in the given initialisation.
The system converts potential energy into kinetic energy and vice versa
to equilibrate the system. 
This conversion of energy was already addressed in the previous subsection 
(particle production). The initialisations of IQMD and BQMD 
yield different pion numbers due to different balances of kinetic and 
potential energy.  

The counterbalancing parts are the kinetic
pressure which causes an expansion and the density dependence of the
potentials which may cause attraction for low densities and repulsion
for high densities.

For this it may be interesting to regard the mean density
of the system which is the mean value of the density of each particle
averaged over all particles. It should be noted that this value may be
sensible to density fluctuations in the center which do only slightly effect
the root mean square radius.

A time evolution of single nuclei % during a period of about 60 fm/c 
 yields for the Au case (hard eos) mean densities
(density per particle averaged over all particles) changing between
about $\rho=0.14$ and 0.15 $\rm fm^{-3}$ for IQMD and between
$\rho=0.11$ and 0.18 $\rm fm^{-3}$ for BQMD.
%The propagation time of 60 fm/c corresponds to about the time period
%a heavy nucleus (like Au)  with low incident energy (about 40 AMeV)
% has to pass before his first
% reaction in a peripheral collision (for about $b \approx 12$fm).  
For smaller systems (on a  time period of about 60 fm/c) the
stability of IQMD gets smaller, to e.g. a range of 
$\rho=0.14$ to 0.17 $\rm fm^{-3}$ for a Nb nucleus in IQMD, 
$\rho=0.16$ to 0.19 $\rm fm^{-3}$ for a Ca nucleus in IQMD and 
$\rho=0.16$ to 0.21 $\rm fm^{-3}$ for a Ne nucleus in IQMD, 
while in BQMD the fluctuations remain constant  
($\rho=0.1 - 0.17\, \rm fm^{-3}$) for all three regarded systems.
It should also be noted that these fluctuations increase in IQMD if a
soft eos is used (about $\rho=0.11 -0.15$ for the Au case) and 
decrease in BQMD (to about the same values).

Let us now examine the density profile of a single Au nucleus. 
Fig. \ref{hipl4rho} displays the time evolution of the density profile  
within BQMD. We observe a  
change in the center as well as at the surface. In a considerably large volume
around the center ($r\le 5$ fm) the change of density with time induces 
changes of the compressional energy in a heavy ion reaction. The weakening
of the surface causes an increase of the rms radius and it therefore modifies
the total interaction cross section as well as the probability of e.g.\
particle production processes in particular in peripheral collisions.
Without Yukawa forces 
these fluctuations are even larger. The reason for these fluctuations are
the lack of pressure built up by the Fermi momentum when the nucleus
gets compressed. This can be verified by initializing HQMD with
the full Fermi momentum but otherwise as above. 

The time evolution of the density profile in 
IQMD is  displayed in Fig.% \ref{hipl7rho} 
\ref{hipl6rho} for a gold nucleus with % and for two Gaussian widths,
the default width of $L=8.66\,fm^2$.
We find strong fluctuations at $r=0$ but a stable shape at the surface and
a rather stable rms radius.
We also find that the shape of the $r^2\rho(r)$ distribution 
shows better stability for the $L=8.66\,fm^2$ case than for $L=4.33\,fm^2$. 
This also motivated the choice of $L$ in IQMD.

It has now to be tested whether these fluctuations of the density profile 
cause uncertainties in the determination of observables. This is tested 
by changing the initialisation distance $d$ 
(with respect to the minimum distance for a head-on
collision) which is defined as ($R$ are the radii of the nuclei):
$$d= z(\hbox{center of proj.})- z(\hbox{center of target}) - R(\hbox{proj.})
- R(\hbox{targ.}) $$
By changing this distance we allow the nuclei to change their profile 
according to the internal forces before they come into nuclear contact. 
For the ideal case this technical parameter
should have no influence on the observables. In reality
the observables depend on it, however, in most cases weakly.

A strong influence of the density fluctuations on observables 
has been found particularly within BQMD for the collective sideward flow. 
Depending on the initially chosen 
distance the total directed transverse momentum transfer in the 
reaction Au (1 GeV/nucleon, b=3fm) + Au varies strongly as
it can be seen in fig,\ref{hipbpx5}. Although the absolute magnitude of the flow depends on 
whether a Yukawa interaction is employed or not, this variance is observed
in both cases. IQMD, however, run with the default parameters shows a
much weaker dependence on this technical parameter. 
In the calculation the flow varies as a function on the
initialisation distance only by about 10\%. 
An IQMD calculation with  $L=4.33\,fm^2$ shows less stability. 
The flow values are decreasing with initialisation distance. This corresponds
to the effect that the density of maximum
stability ($\rho=0.15 \rm \, fm^{-3}$ for $L=4.33 \, fm^2$) is not
equal to the initialisation density  $\rho_{ini}=0.17\, \rm fm^{-3}$.
This was originally the
motivation for the use of a system dependent width in IQMD.
A BQMD calculation with $L=8.66$, however, still shows strong 
fluctuations. The fluctuations in BQMD decrease if the Fermi momentum 
of the initialized nuclei is increased. However, in this case the binding energy
of the nucleus becomes smaller and spurious particle evaporation may be
effected. 

For the observables for Au(1GeV)+Au 
discussed in this paper we find the following maximum
deviation (in a range of initialisation distances between 0 and 
$d_{max}=13$ fm)
from the default values: for the IQMD initialisation 
10\% concerning flow and kaon multiplicity 
and 8\% concerning  pion multiplicity
and for the BQMD initialisation 70\% concerning flow, 45\% concerning kaon and
 35\% concerning pion multiplicity.
For the fragmentation of Au(150Mev)+Au b=3fm both models yield (in their
default modes) about 8-15\% deviation in the number of IMFs ( a rise from 6.6
to about 7.1 for IQMD and a fall from 12.7 to about 10.8 for BQMD).

It should be noted that the reported errors also include about 5-10\%
statistical fluctuations and that for initialisation distances larger than
13 fm the deviations may still increase for some calculations. The value
of 13 fm has been chosen since it is the difference in the effective
distance to the first reaction point between a central and a very peripheral 
collisions.

In any case, these fluctuations cause
an additional systematic error which has to be added to the statistical one
if one compares with data. Because these fluctuations are stronger
using a BQMD initialisation the systematical error is larger there.

In conclusion we find that both models show fluctuations of the density
profile at $r=0$. However IQMD shows in its default mode a rather stable
$r^2\rho(r)$ shape while for BQMD the maximum of $r^2\rho(r)$ changes in
time. This yield artificial vibration modes which influence the stability of the
nucleus and therefore cause systematic errors which in most cases are
stronger in BQMD than in IQMD. Especially  the discussion of dynamical
variables like flow and particle production within BQMD has to be regarded
very cautiously.

%%%%%%%%%%%%%%%%%%%%%%%%%%%%%%%%%%%%%%%%%%%%%%%%%%%%%%%%%%%%%%%%%%%%%%%%%%
\section{Summary and concluding remarks}
We have compared different realizations of the Quantum Molecular
Dynamics model.
The different realizations differ in certain input variables as they are
comparatively presented in table \ref{qmdinputs}. Some of these parameters
are purely technical, some are physical. The latter parameters are 
constrained by experimental observations but are not completely fixed. 
If the same parameters are employed the
result of the different programs are identical in between the error.
This is a remarkable achievement in view of the several thousand program
lines of each of these programs.

The HQMD realization allows for the first time to compare in detail the
influence of different inputs on the different observables. The most
important input is the choice of the Gaussian width $L$ or in other words
the interaction range of the nuclear potential. A change of the interaction 
range causes differences in the density 
profile of a ground state nucleus and in the strength  of the density 
gradient. Thus a smaller interaction range
yields an enhanced flow, enhanced fragment production and (which was not shown)
a reduced numbers of pions (slight changes)  and kaons (larger changes). 
The interaction range determines the surface properties of the nuclei as
well as their binding energy. Only in infinite
nuclear matter the binding energy is independent
of this quantity. The experimental value of these observables allow to 
fix the range of possible values. 
 
The choice of the cross section employed in HQMD, BQMD or IQMD yield 
slight changes in the flow (and also slight differences in the stopping) but 
has no influences on the fragmentation. Pion production in HQMD and IQMD
are comparable.

An important factor is the choice of the ground state description. This 
choice effects the results in flow and particle production, but does not 
influence the fragmentation pattern. The flow at high energies is found to
be stronger if the initialisation of the nucleus is more compact.
At low energies these differences vanish, instead the average Fermi momentum
becomes important. The Fermi momentum corresponding to (infinite) nuclear
matter are necessary to stabilize the nucleus against artificial vibrations 
and yield better agreement with experimental flow data. 
At the same time a large Fermi momentum lowers the binding energy.
The binding energy of heavy nuclei is reproduced if local Fermi
momenta are employed.  

As a conclusion we find that there are variables which are very robust 
against a change of the technical or physical parameters, e.g.\ the
rapidity distribution. Others like the fragmentation pattern depend 
on the range of the interaction only. Other observables, like the directed
flow have a very strong dependence on many details of the calculation
and slight differences between the different QMD flavours yield
large differences in this observable.  

In the BQMD proper binding energies have been achieved on the cost of
large dynamical fluctuations in the initial state and by a moderate energy
non-conservation ($\approx$ 2 MeV/nucleon for single nuclei)
on an event by event basis.
The IQMD-model offers rather stable density distributions and good
energy conservation, however for the price of nucleon evaporation and
and improper binding energies ($E_{bind}\approx 4-5$ MeV/nucleon for heavy
nuclei instead of 8 MeV/nucleon).

The choice between a parametrization which yields the proper ground
state energy of projectile and target and that which yield 
the necessary Fermi momentum
to obtain the observed flow is not satisfying. Therefore
work is in progress to modify the bare interaction between the nucleons
inside the nucleus in a way which allows to obtain both at the same time.
In particular one has to account for the peculiar dependencies between the
various parts of the model. Initialisation, propagation, hard collisions,
and Pauli-blocking cannot be treated independently from each other.
Forces and cross-sections are connected. The specific choice of the 
saturation density, and thus also that of the central density of heavy nuclei
influences the physical output.
These aspects 
are part of the effort to obtain a new unified 
QMD scheme which covers the energy range between 25 MeV/nucleon and 
200 GeV/nucleon. The aim of this new model will be to cover the
best possibilities for a reliable description of the different
aspects of heavy ion collisions \cite{uqmd}.

\subsection*{Acknowledgements}
This work was in part supported by the French Institut National de Physique
Nucl\'eaire et de Physique des Particules (IN2P3), by the German 
Bundesministerium f\"ur Bildung und Forschung  (BMBF), by the 
Deutsche Forschungsgemeinschaft (DFG) and by the Gesellschaft f\"ur 
Schwerionenforschung (GSI).

%%%%%%%%%%%%%%%%%%%%%%%%%%%%%%%%%%%%%%%%%%%%%%%%%%%%%%%%%%%%%%%%%%%%%%%%%%
%\bibliographystyle{unsrt}
%\begin{thebibliography}{10}

\pagebreak

\begin{table}
\begin{tabular}{lccccc}
 &$\alpha$ (MeV)  &$\beta$ (MeV) & $\gamma$ & $\delta$ (MeV) &$\varepsilon \, 
 \left(\frac{c^2}{\mbox{GeV}^2}\right) \!\!\!\!$ \\
\hline
 S  & -356 & 303 & 1.17 & ---  & ---    \\
 SM & -390 & 320 & 1.14 & 1.57 & 500  \\
 H  & -124 & 71  & 2.00 & ---  & ---    \\
 HM & -130 & 59  & 2.09 & 1.57 & 500  \\
\end{tabular}
\caption{\label{eostab} Parameter sets for the nuclear equation of 
state used in the
QMD model. S and H refer to  the soft and hard equations of state, 
M refers to the 
inclusion of momentum dependent interaction.}
\end{table}

\begin{table}
\begin{tabular}{ccc}
$x=\sqrt{s}$ (GeV) & $a$ (fm) & $b$ \\\hline
2.104 -- 2.12 	&  $294.6 \; (x - 2.014)^{2.578}$ 
	& $19.71\; (x - 2.014)^{1.551}$   \\
2.12 -- 2.43	& $\frac{0.01224}{(x- 2.225)^2 + 0.004112}$  
	& $19.71\; (x - 2.014)^{1.551}$ \\
2.43 -- 4.50 	& $(2.343/x)^{43.17}$ & $33.41 \; 
\arctan( 0.5404 \;(x-2.146)^{0.9784})$  \\
\end{tabular}
\caption{ \label{inelast} $a(s)$ and $b(s)$ as functions of the c.m. energy. }
\end{table}

\begin{table}
\begin{tabular}{lcc}
$x=\vert \Delta \vec{p} (CM)\vert/1 $ GeV & $\sigma_{el}$(mb) for pp, nn &
$\sigma_{el}$ (mb) for pn \\ \hline
$x<0.8$ & $ 23.5 + 1000\cdot (0.7-x)^4 $ & $33 + 196\cdot |0.95 -x|^{2.5} $ \\
$0.8 < x < 2$ & $1250/(x+50) \quad - \quad 4\cdot (x-1.3)^2 $ & $31/\sqrt{x}$ \\
$2 < x$ & $ 77/(x+1.5)$ & $77/(x+1.5) $ \\
\end{tabular}

\caption{\label{sigcugnew} 
Elastic cross section parametrization used in QMDRKNC and HQMD as a function
of the relative momentum in the CM-frame}
\end{table}

\begin{table}
\begin{minipage}{\textwidth}
\qtabstart
\qtab{Input}{BQMD}{IQMD}{HQMD}
\qtab{Initialisation}{Wood-Saxon}{hard sphere}{both}
\qtab{Init. distance}{3fm}{0fm}{3fm}
\qtab{Gaussian width}{4.33 fm$^2$}{8.66 fm$^2$}{4.33 fm$^2$}
\qtab{Coulomb forces}{$Z_p =Z_n$}{$Z_p=1, Z_n=0$}{$Z_p=Z_n$}
\qtab{Yukawa forces}{L=1.5 fm}{L=0.4 fm}{none}
\qtab{Yukawa adjust.}{$t_1(i)=t_1 -\sum_j U_{ij}^{Yuk}/\rho$}{
$t_1=t_1 - \kappa*V_0^{Yuk}*L^{Yuk}/L^{3/2}$}{none}
\qtab{actio = reactio}{on the average}{`exact'}{`exact'}
\qtab{asymmetrie forces}{none}{$t_6/\rho_0\, T_3^i T_3^j\, \delta(\vec{r_i} -
\vec{r_j})$}{none}
\qtab{forces on $\pi$}{no $\pi$}{Coulomb}{no force}
\qtab{cross sections}{Cugnon 1981}{VUU 1986}{Cugnon 1989}
\qtab{particles}{$N, \Delta$}{$N,\Delta,\pi$}{$N, \Delta, N^*,\pi$}
\qtabstop
\end{minipage}
\caption{ \label{qmdinputs}
Comparison of the different realizations BQMD, HQMD and IQMD concerning
the different ingredients of the inputs.
}
\end{table}

\pagebreak
\begin{figure}
\caption{\label{hipsigma}
The elastic and inelastic cross sections for proton-proton (pp) and 
proton-neutron (pn) used in IQMD. The neutron-neutron cross section 
is assumed to equal to the pp case.
The total cross section is equal to the sum of elastic and inelastic
cross section.}
\end{figure}

\begin{figure}
\caption{ \label{vuu0dndy}
Rapidity distributions $dN/dy$ of nucleons in the reaction
Au(1 GeV/nucleon)+Au, b=3fm for VUU, IQMD and BQMD. In all calculations
a hard equation of
state without momentum dependent interactions has been used.}
\end{figure}

\begin{figure}
\caption{ \label{vuu0pxe}
Excitation function of the system Au+Au at $b=3$fm impact parameter obtained
with BQMD, IQMD and VUU in their default versions. 
}
\end{figure}

\begin{figure}
\caption{ \label{hipl4px}
Excitation function of the system Au+Au at $b=3$fm impact parameter obtained
with IQMD using the width of $L=4.33\,fm^2$  and $L=8.66\,fm^2$.
}
\end{figure}

\begin{figure}
\caption{ \label{hip11pxe}
Excitation function of the system Au+Au at $b=3$fm impact parameter obtained
with HQMD (default collision term) using the BQMD initialisation, a hard
sphere initialisation with reduced Fermi momentum and with full Fermi 
momentum and the IQMD initialisation.
}
\end{figure}

\JKalt{
\begin{figure}
\caption{ \label{hip21pxe}
Excitation function of the system Au+Au at $b=3$fm impact parameter obtained
with HQMD (BQMD-initialisation)
using the HQMD default collision terms and the collision terms of BQMD and 
IQMD.
%\\ \bf Can also be done with IQMD-ini, but data have to be restored after
%disk-sweep on GSI disks. \rm
}

\end{figure}
}

\begin{figure}
\caption{ \label{vuu0pxy}
Comparison of the transverse flow $p_x(y)$ of nucleons in the reaction
Au(1AGeV)+Au b=3fm for VUU, IQMD and BQMD, all using a hard equation of
state without momentum dependent interactions.}
\end{figure}

\begin{figure}
\caption{ \label{hip15a7}
Fragment mass distribution obtained by BQMD and IQMD for the 
system Au(150MeV) +Au  b=3fm, both  with $L=4.33\,fm^2$ and $L=8.66\,fm^2$.
}
\end{figure}

\begin{figure}
\caption{ \label{hip7dyp}
Rapidity distributions of pions
for  Au(1 GeV/nucleon)+Au reactions at b=3fm obtained with QMD 
employing the BQMD initialisation (circles) and 
the IQMD initialisation (diamonds).}
\end{figure}

\begin{figure}
\caption{ \label{hiptrms4}
Time evolution of the root mean square radii of a single Au nucleus in
coordinate and momentum space obtained with IQMD and BQMD using a hard and
a soft eos.
}
\end{figure}

\begin{figure}
\caption{ \label{hipl4rho}
Time evolution of the
density profiles $\rho(r)$ obtained for a
Au nucleus initialized with  BQMD (with Yukawa) using the width of $L=4.33\,fm^2$.
}
\end{figure}
\begin{figure}
\caption{ \label{hipl6rho}
Time evolution of the
density profiles $\rho(r)$ obtained for a
Au nucleus initialized with  IQMD using the width of $L=8.66\,fm^2$.
}

\begin{figure}
\caption{ \label{hipbpx5}
Dependence of the flow obtained from the system Au(1 AGeV)+Au at b=3fm
on the initialisation distance using IQMD with $L=8.66\,fm^2$ and BQMD with
$L=4.33\,fm^2$.
}
\end{figure}

\end{figure}

\pagebreak
\epsfxsize=12cm
\newcommand{\EPS}[2]{
\begin{center} \epsffile{#1} \nopagebreak \mbox{\Large\bf 
Figure {\ref{#2}} } \end{center}}
%\begin{center} \psfig{figure=#1}\end{center}}
\EPS{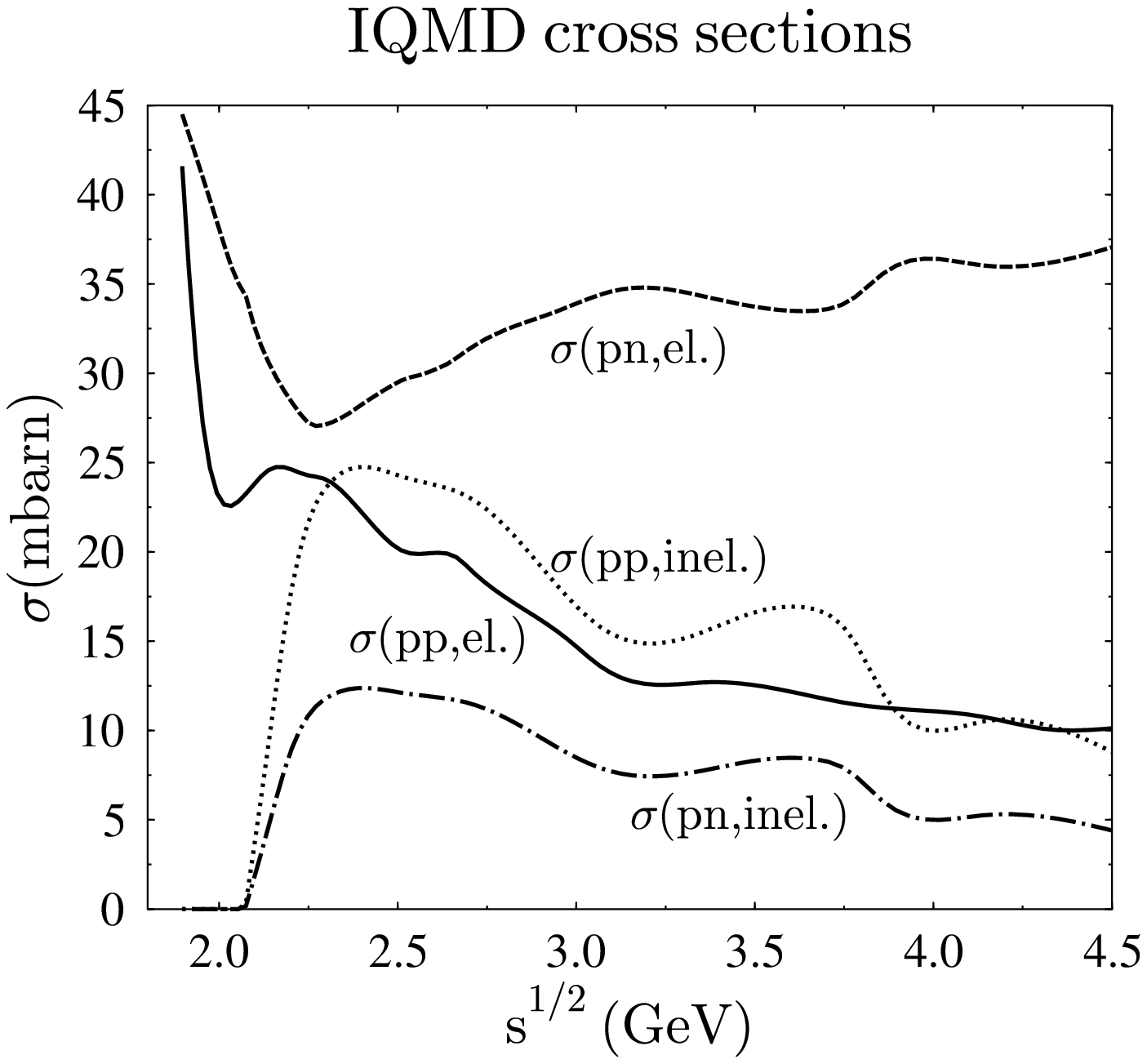}{hipsigma}
\epsfxsize=12cm
\EPS{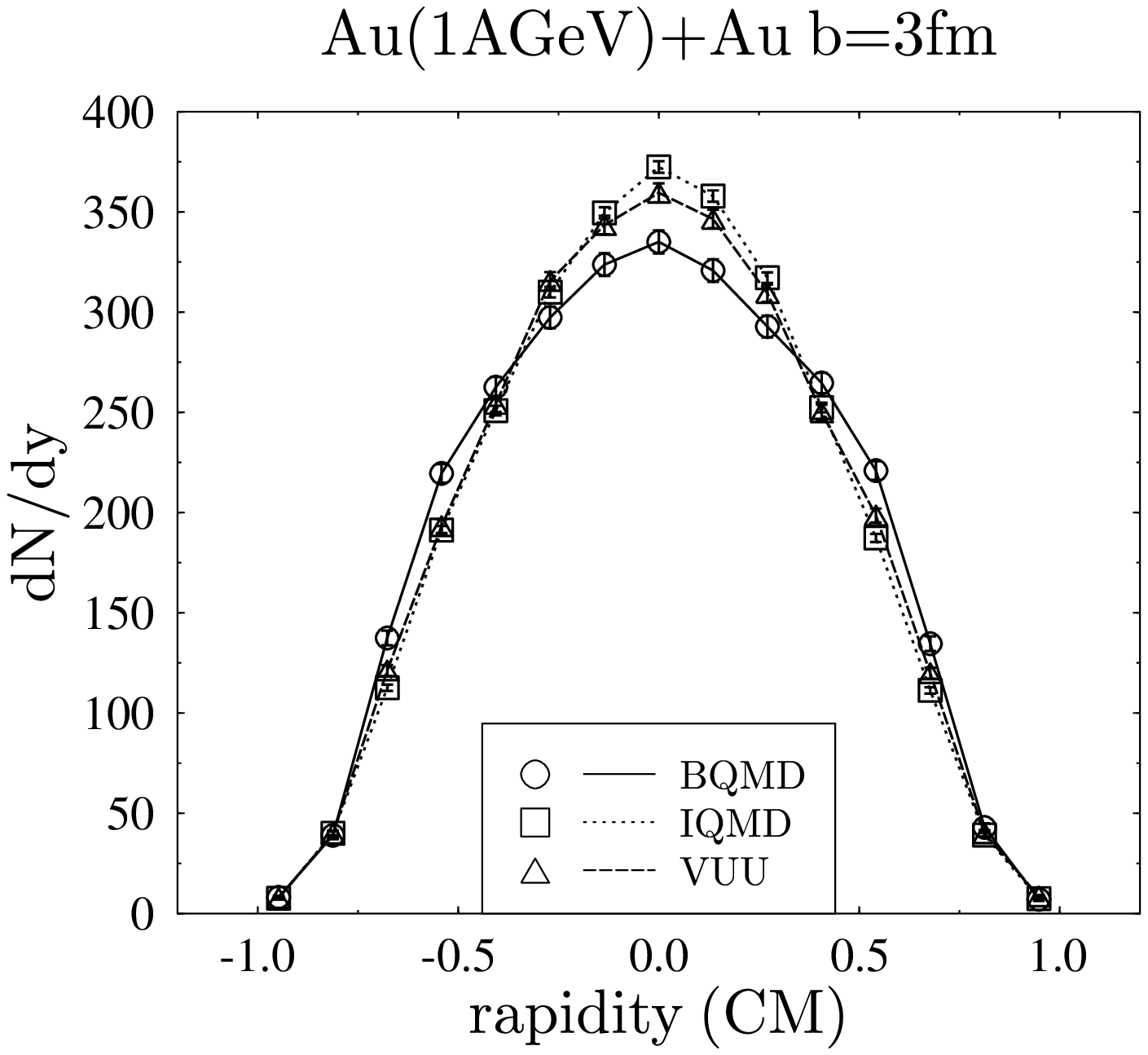}{vuu0dndy}

\epsfxsize=12cm
\EPS{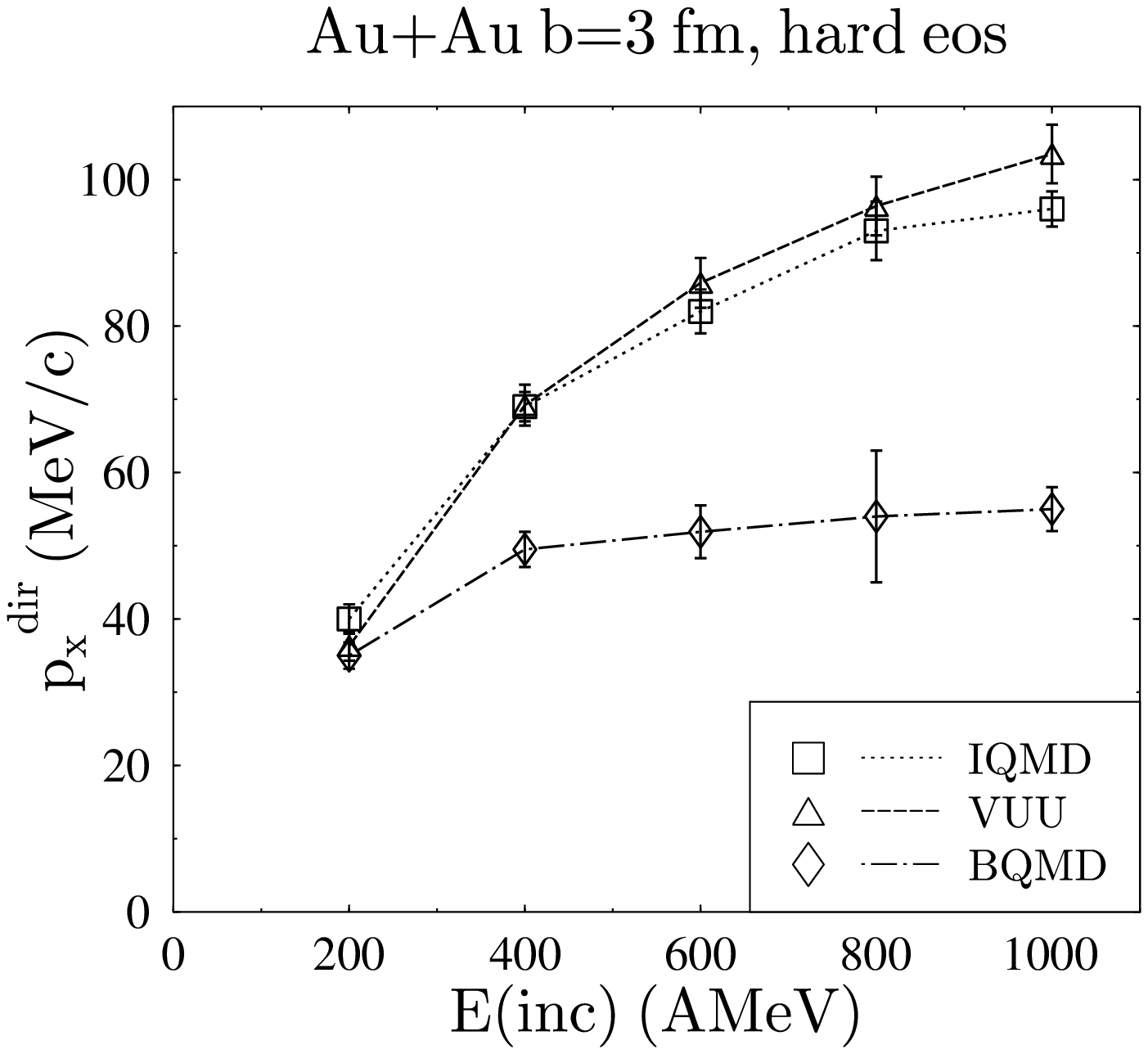}{vuu0pxe}

\epsfxsize=12cm
\EPS{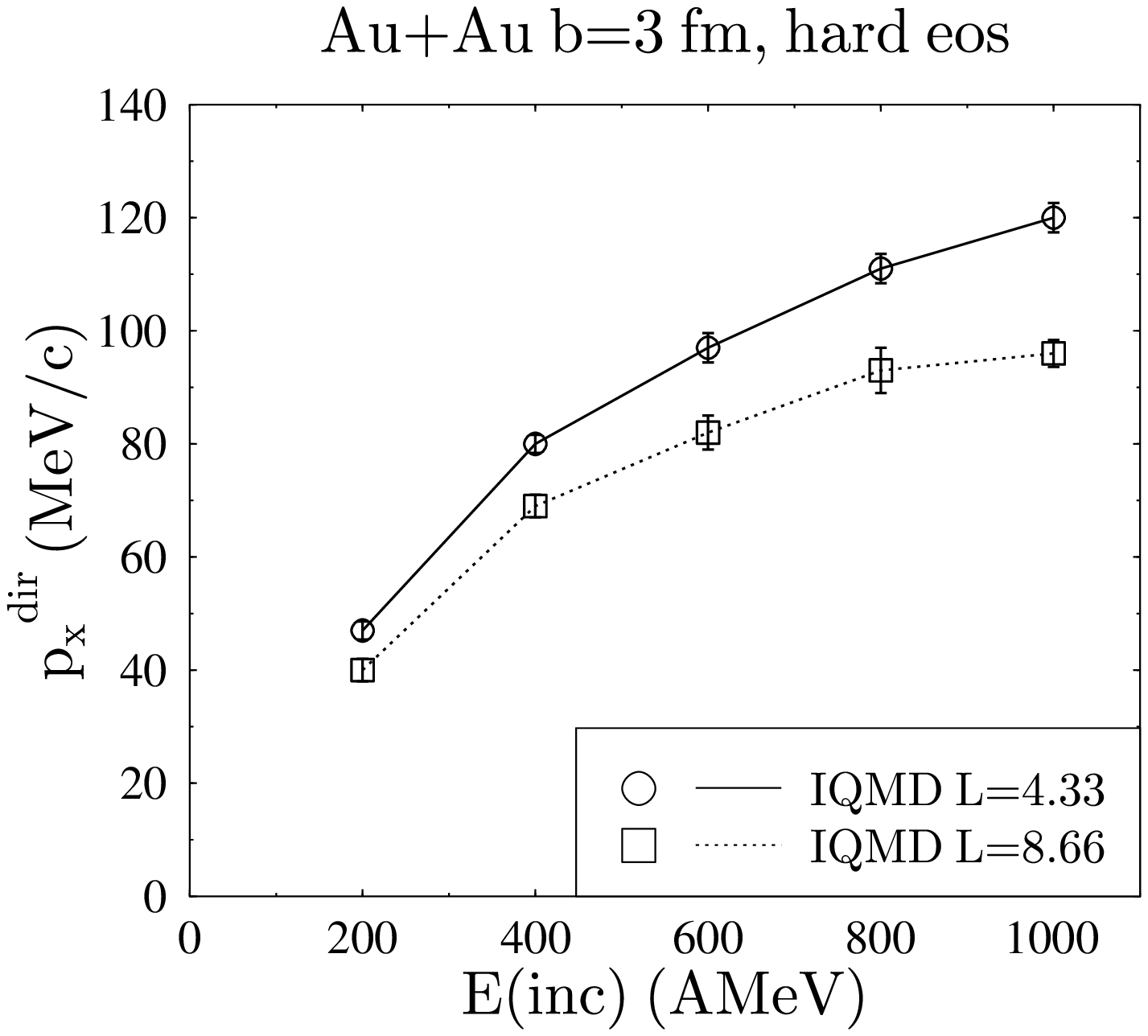}{hipl4px}
\epsfxsize=12cm
\EPS{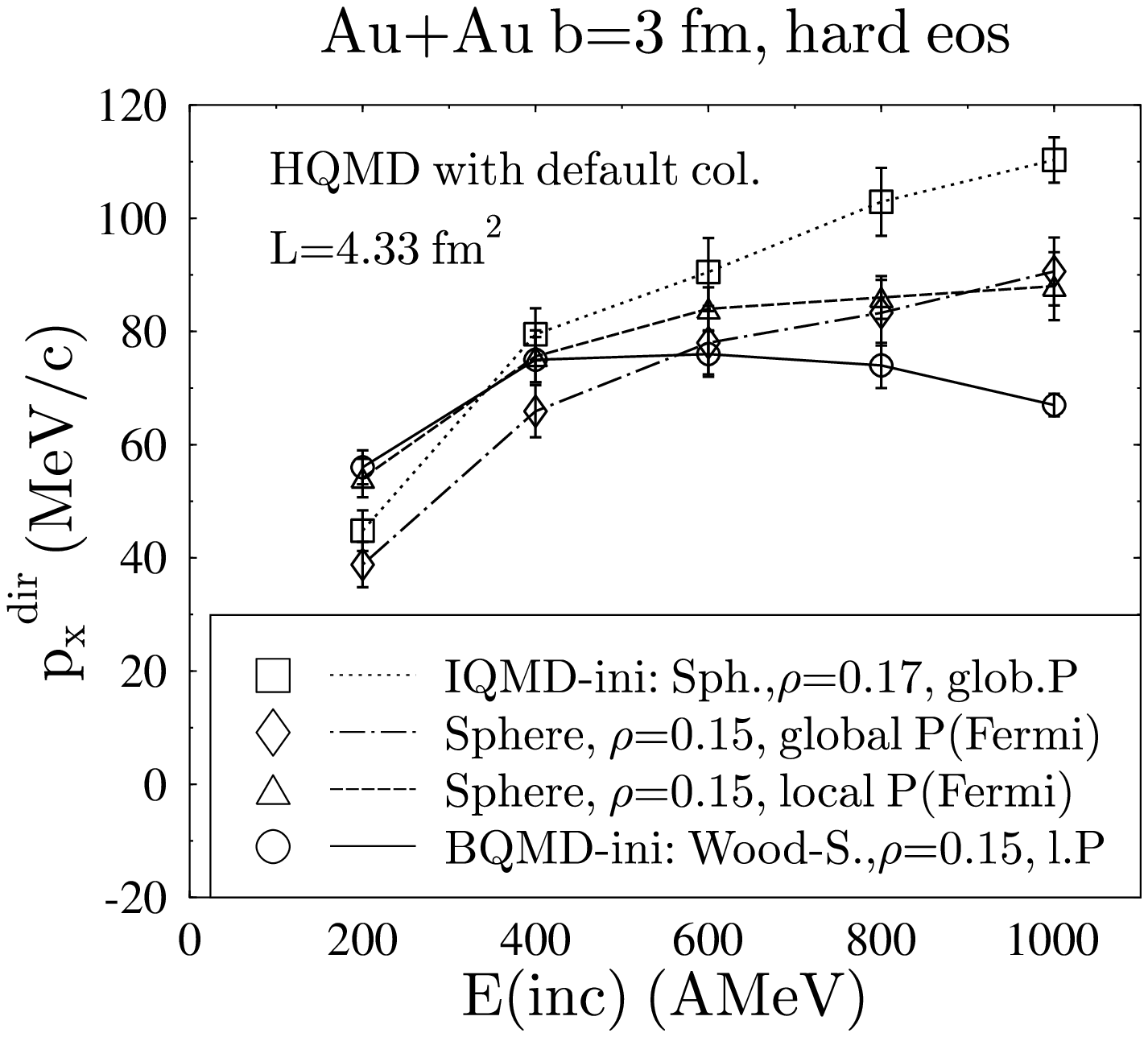}{hip11pxe}

%%\EPS{hip21pxe.eps}

\epsfxsize=12cm
\EPS{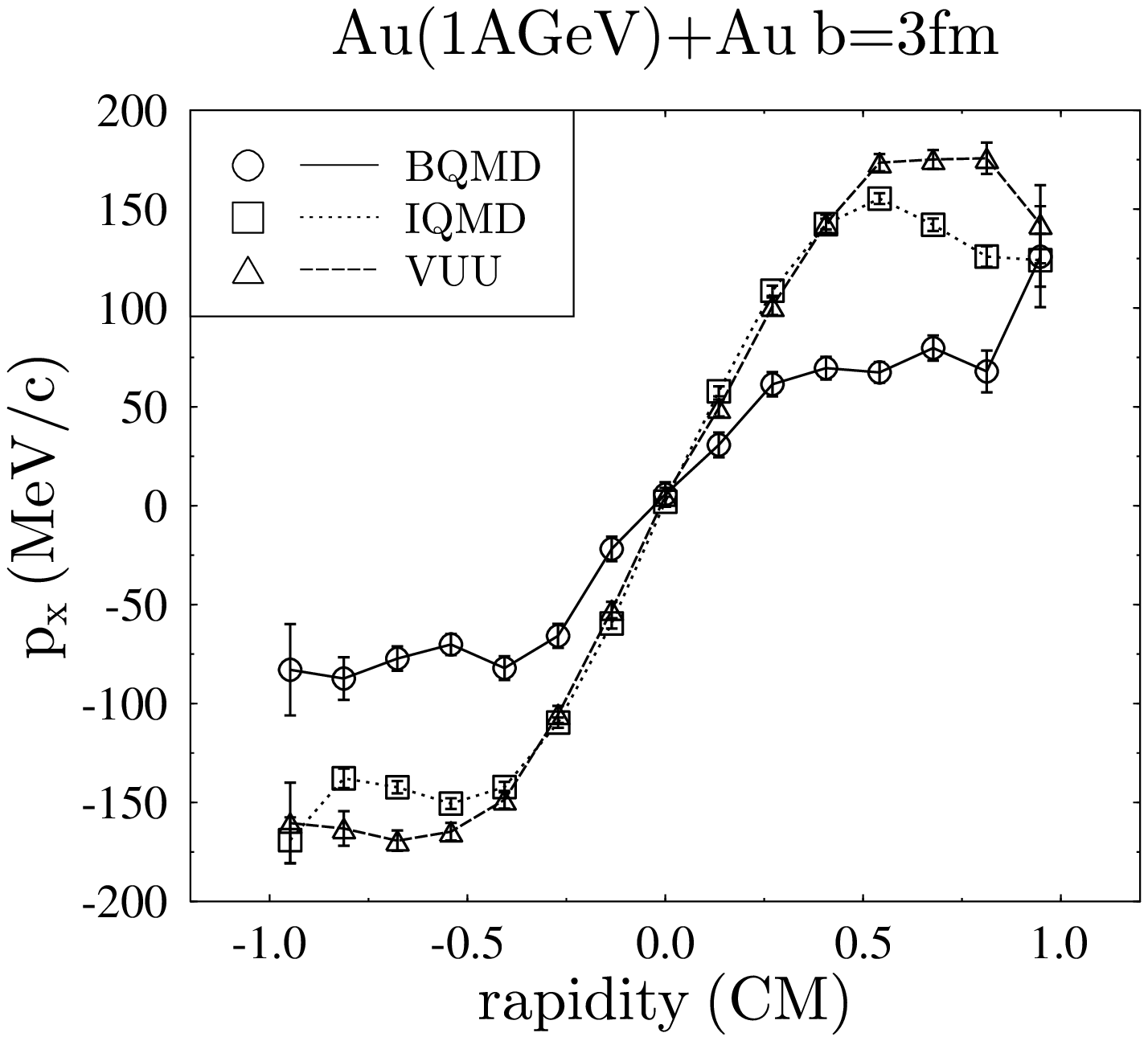}{vuu0pxy}
\epsfxsize=12cm
\EPS{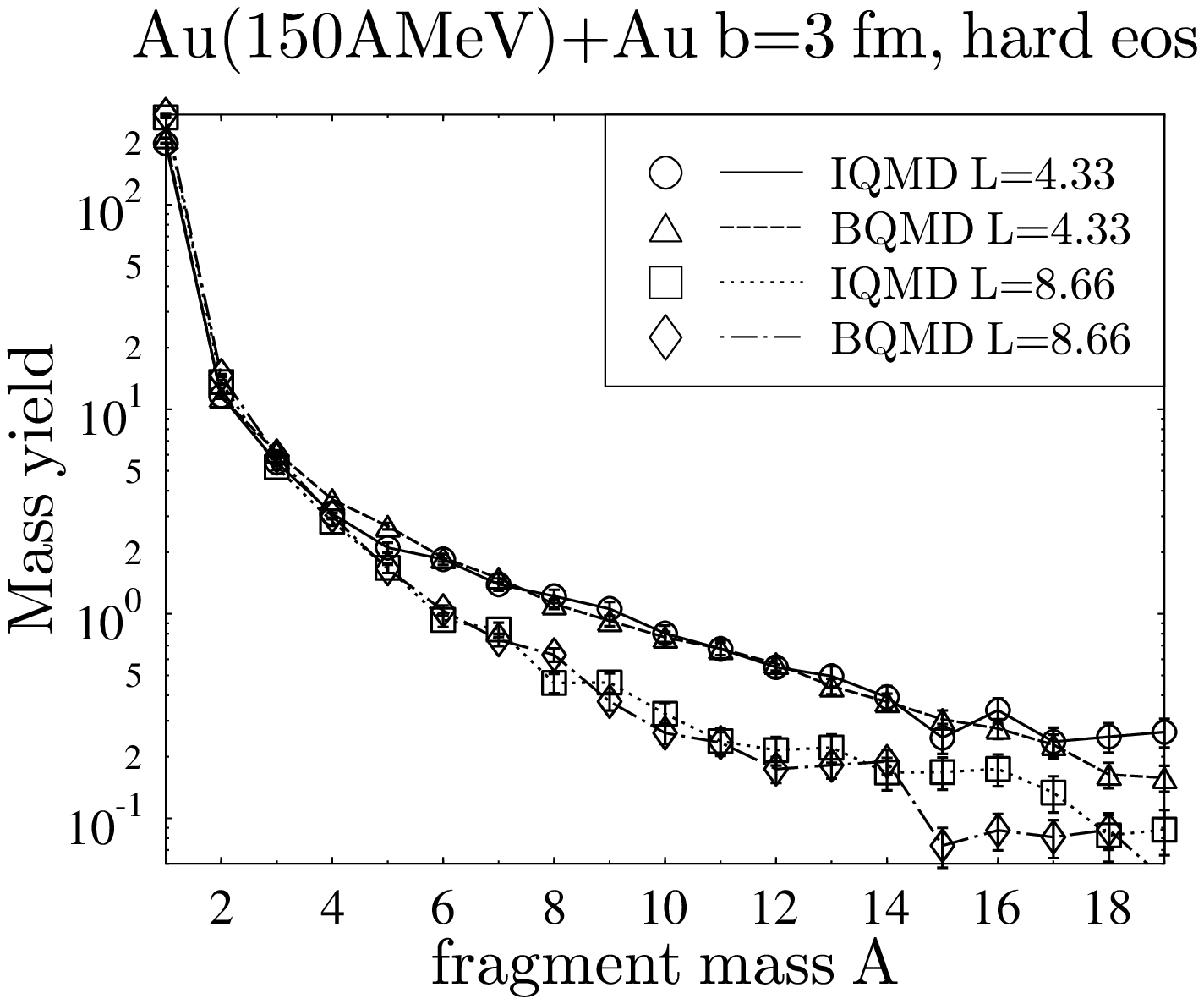}{hip15a7}
\epsfxsize=12cm
\EPS{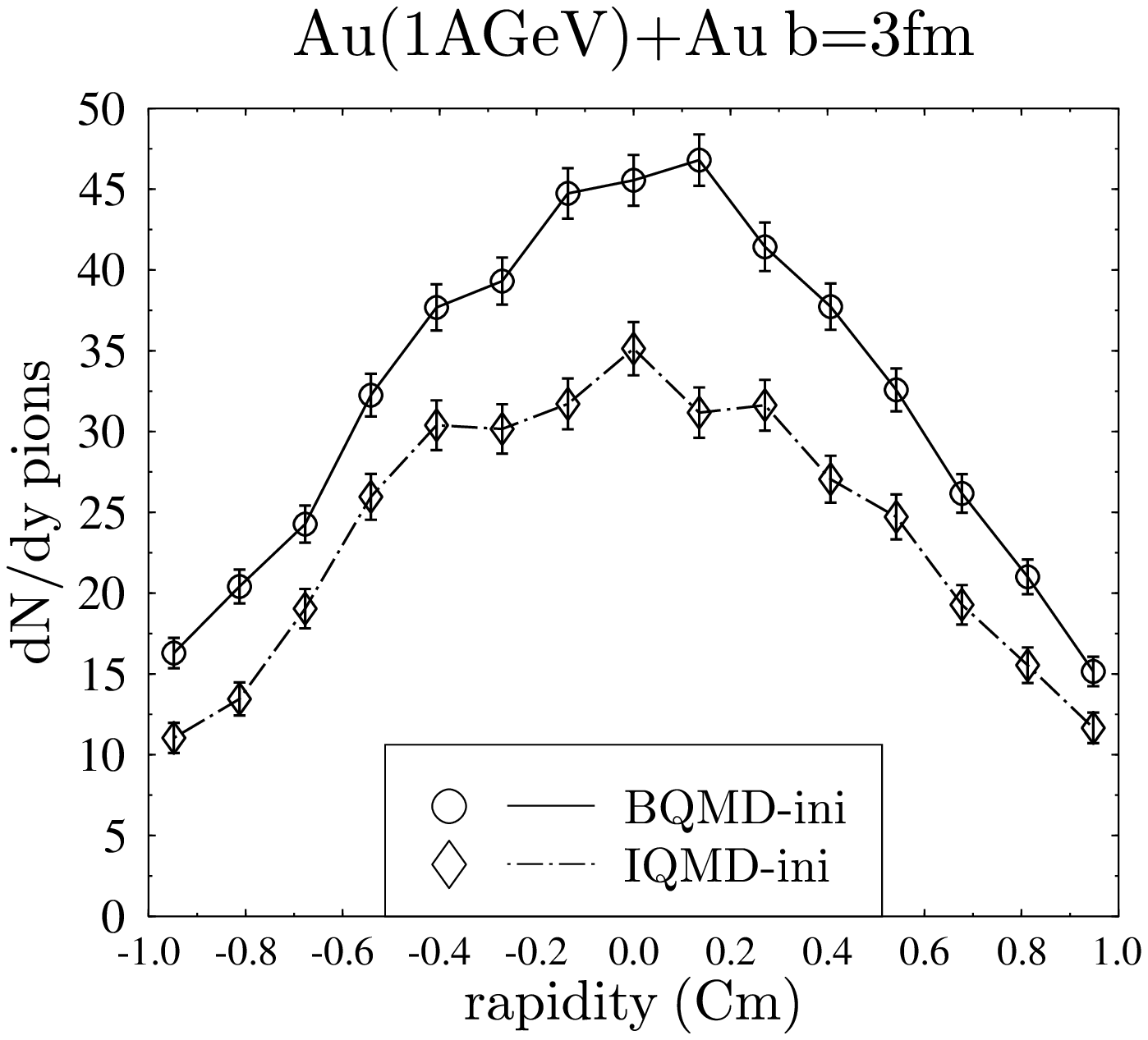}{hip7dyp}
\epsfxsize=12cm
\EPS{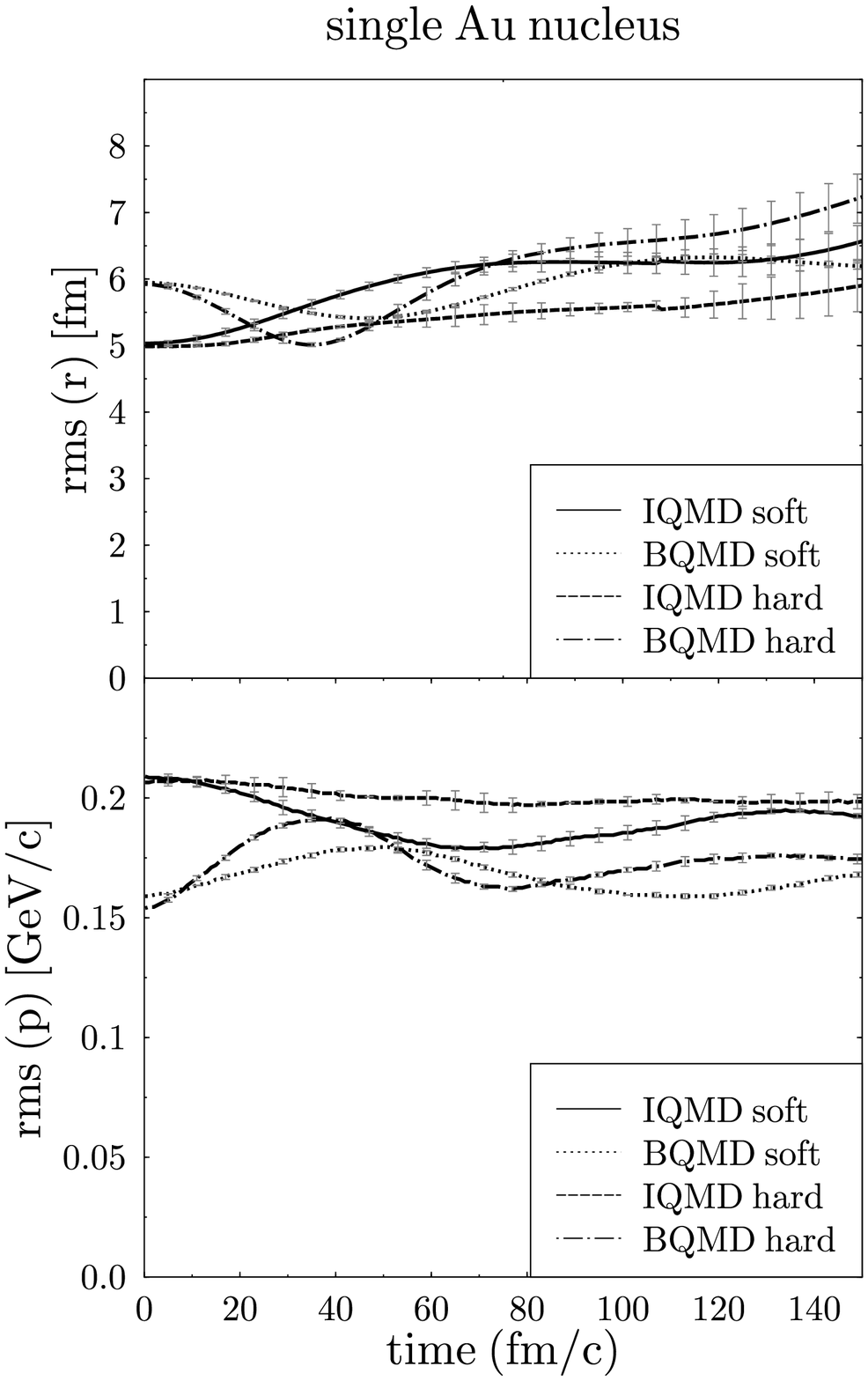}{hiptrms4}

\epsfxsize=12cm
\EPS{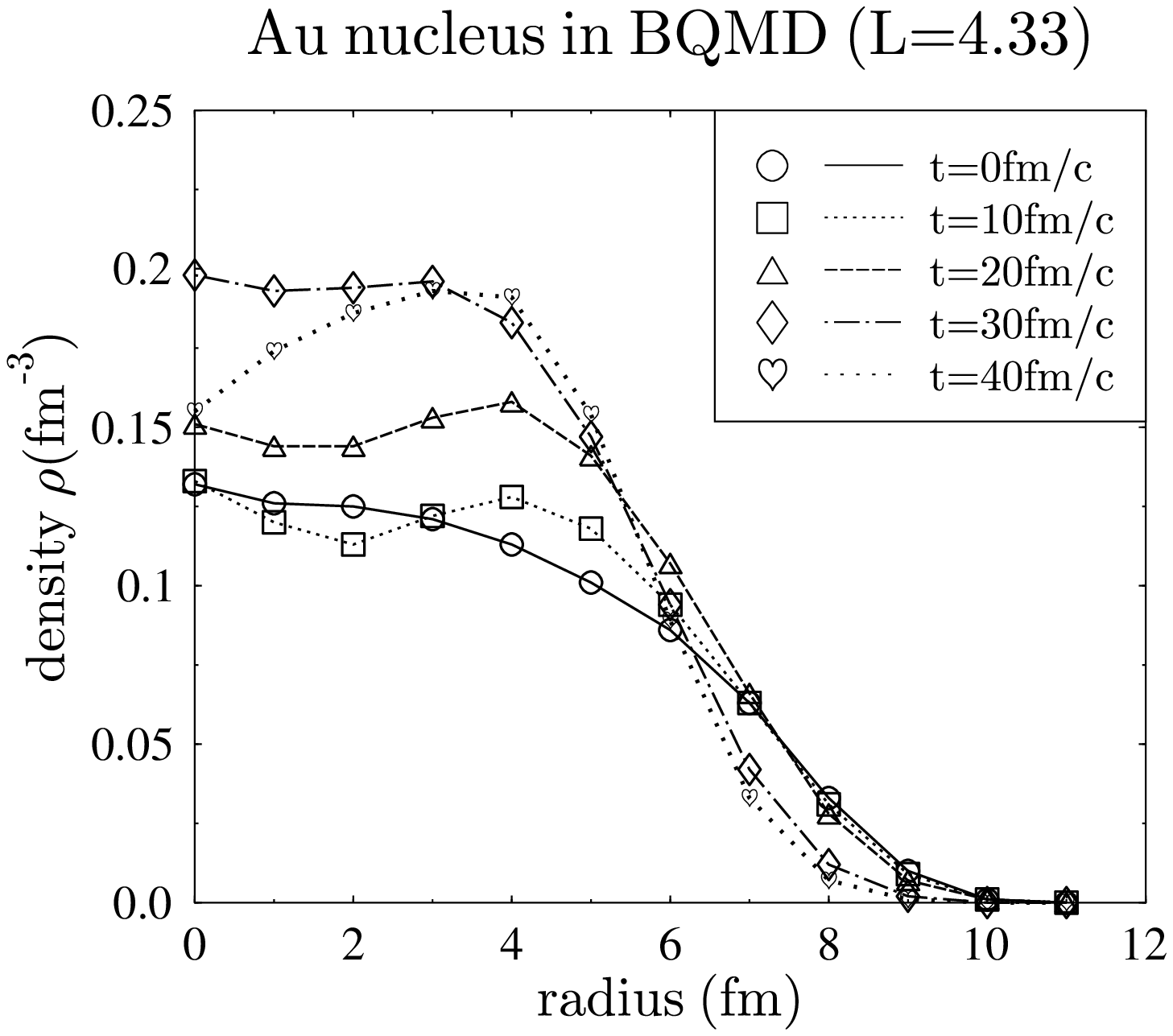}{hipl4rho}
\epsfxsize=12cm
\EPS{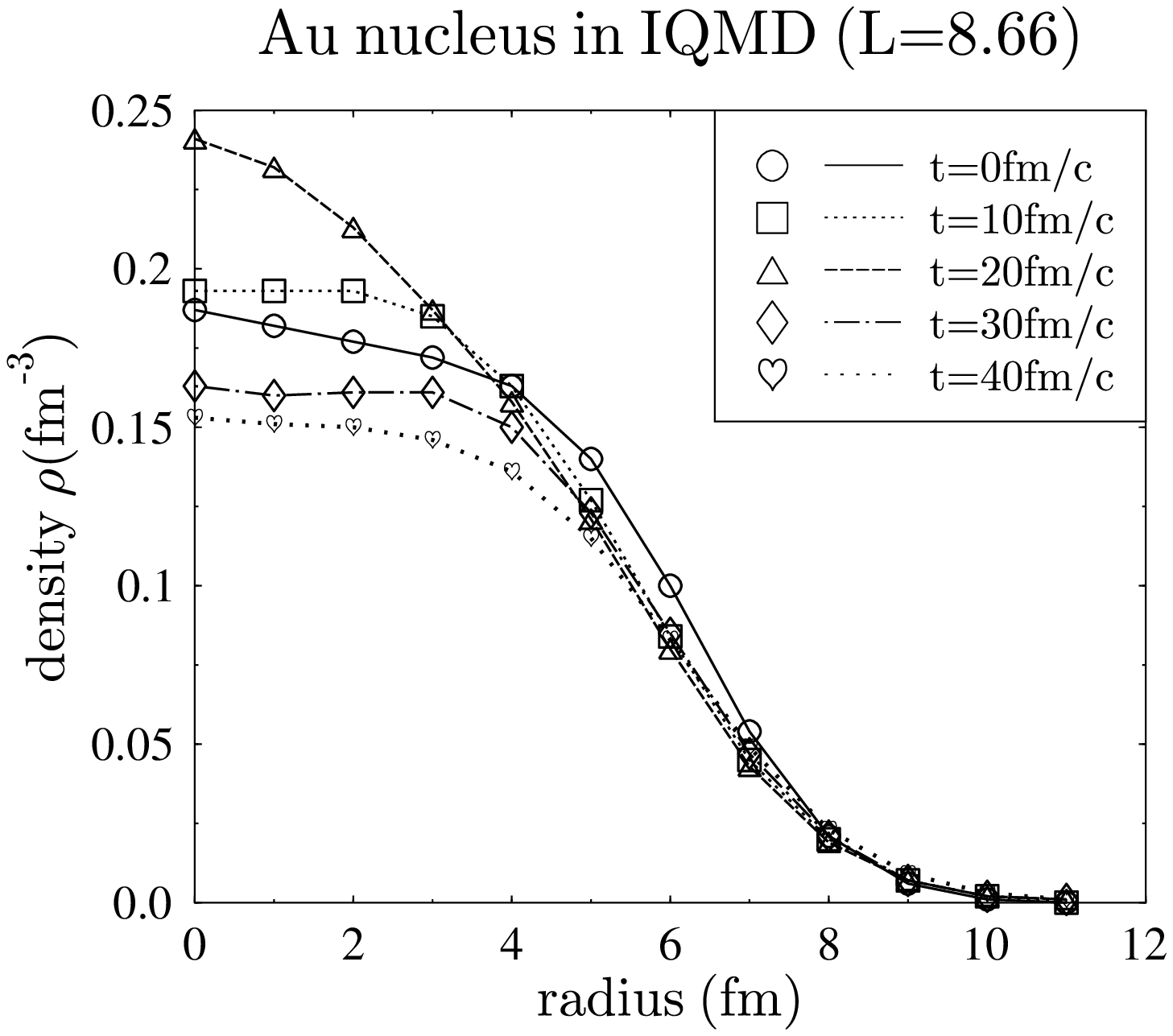}{hipl6rho}

\epsfxsize=12cm
\EPS{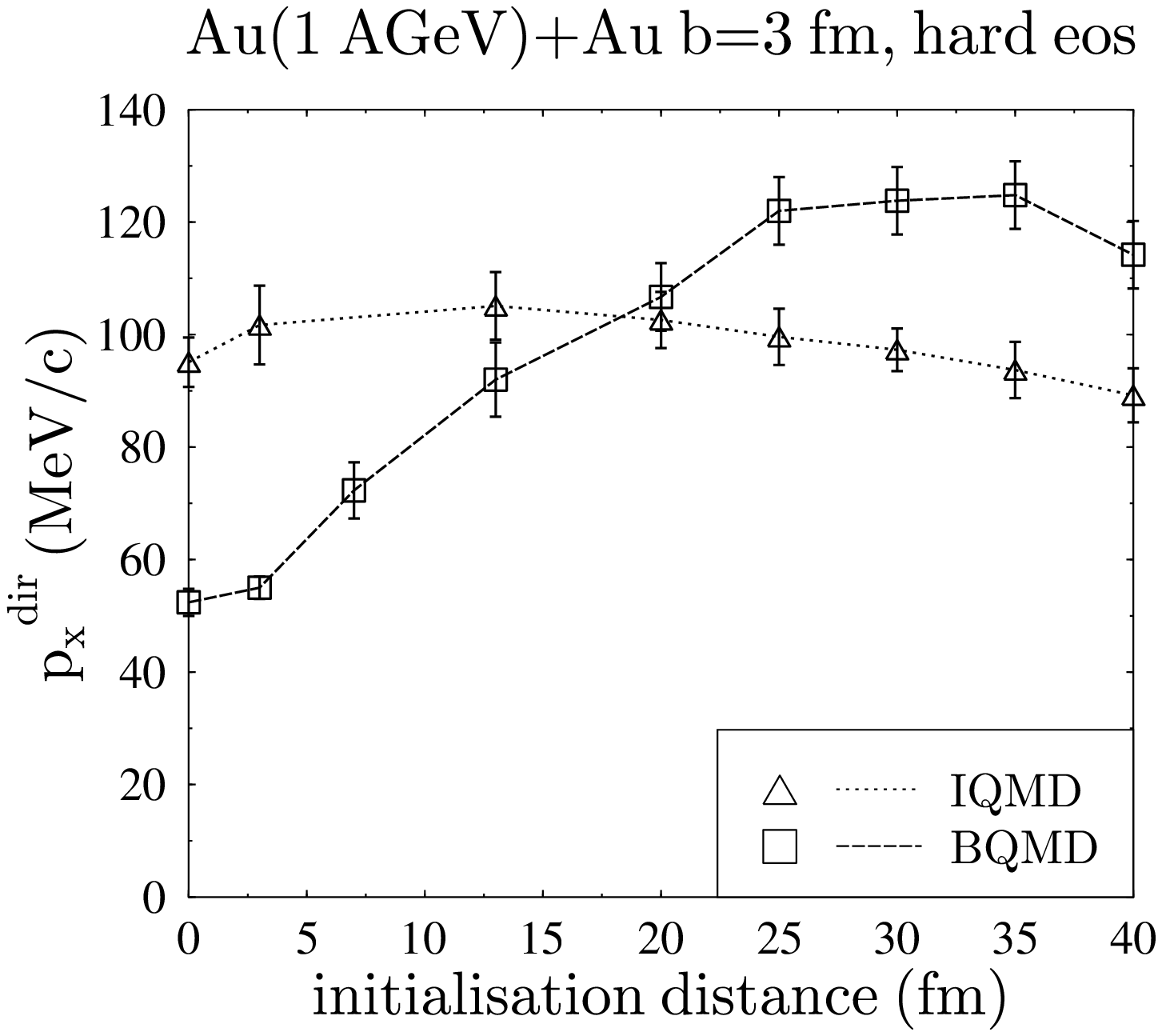}{hipbpx5}


\begin{references}

\bibitem{sch68}
W.~Scheid, R.~Ligensa, and W.~Greiner.
\newblock Phys.~Rev.~Lett. {\bf 21}, 1479 (1968).

\bibitem{cse86}
L.~P.~Csernai and J.~I.~Kapusta.
\newblock Phys.~Reports~{\bf 131}, 225 (1986).

\bibitem{sto86}
R.~Stock.
\newblock Phys.~Reports~{\bf 135}, 261 (1986).

\bibitem{st86}
H.~St\"ocker and W.~Greiner.
\newblock Phys.~Reports~{\bf 137}, 277 (1986).

\bibitem{cl86}
R.~B.~Clare and D.~Strottman.
\newblock Phys.~Reports~{\bf 141}, 179 (1986).

\bibitem{schue87}
B.~Sch\"urmann, W.~Zwermann and R.~Malfliet.
\newblock Phys.~Reports~{\bf 147}, 3 (1986).

\bibitem{cas90}
W.~Cassing, V.~Metag, U.~Mosel and K.~Niita.
\newblock Phys.~Reports~{\bf 188}, 361 (1990).


\bibitem{ai91}
J.~Aichelin.
\newblock  Phys.~Reports~{\bf 202}, 233 (1991).

\bibitem{st80}
H.~St\"ocker, J.~A.~Maruhn and W.~Greiner,
\newblock Phys.~Rev~Lett.~{\bf 44}, 725 (1980).

\bibitem{st82}
H.~St\"ocker, L.~P.~Csernai, G.~Graebner, G.~Buchwald, H.~Kruse, R.~Y.~Cusson,
J.~A. Maruhn and W.~Greiner,
\newblock Phys.~Rev~{\bf C25}, 1873 (1982).


\bibitem{st78}
H.~St\"ocker, W.~Greiner, and W.~Scheid.
Z.~Phys.~{\bf A286}, 121 (1978).

\bibitem{da79}
P.~Danielewicz.
Nucl.~Phys.~{\bf A314}, 465 (1979).

\bibitem{st81}
H. St\"ocker, A.~A.~Ogloblin and W.~Greiner.
\newblock Z.~Phys.~{\bf A303}, 259 (1981).


\bibitem{astro}
S.A. Bludman, Phys. Reports 163 (1988) 47.

\bibitem{flow}
H.A.~Gustafsson et al.,
Phys. Rev. Lett. 52 (1984) 1592.
\\
D.~Beauvis et al.,
Phys. Rev. C27 (1983) 2443;
\\
H.G.~Ritter et al.,
Nucl. Phys. A447 (1985) 3c;
\\
K.G.R.~Doss et al.,
Phys. Rev. Lett. 57 (1986) 302.
\\
H.H.~Gutbrod et al.,
{\rm Phys.~Lett. 216B (1989) 267.}
\\
H.H.\ Gutbrod et al. Rep.\ Prog. Phys. 52 (1989) 1267.

\bibitem{pions}
A.~Sandoval et al.,  
\newblock Phys.~Rev.~Lett. {\bf 45}, 1236 (1980).
\\
R.~Stock et al., 
\newblock Phys.~Rev.~Lett. {\bf 49}, 1236 (1982).
\\
J.~Harris et al.,
\newblock Phys.~Lett. {\bf B153}, 377 (1985).

\bibitem{kru85a}
H.~Kruse, B.~V. Jacak, and H.~St\"ocker.
\newblock  Phys.~Rev.~Lett.~{\bf 54}, 289 (1985).


\bibitem{bert84a}
G.F.~Bertsch, H.~Kruse and S.~Das~Gupta.
\newblock Phys.~Rev.~{\bf C29}, R673 (1984).

\bibitem{ai85a}
J.~Aichelin and G.~Bertsch.
\newblock  Phys.~Rev.~{\bf C31}, 1730 (1985).


\bibitem{greg87}
C.~Gregoire, B.~Remaud, F.~Sebille, L.~Vinet, and Y.~Raffray,
Nucl. Phys. {\bf A465}, 317 (1987).


\bibitem{fmd}
H.~Feldmeier.
\newblock Nucl.~Phys.~{\bf A515}, 147 (1990).

\bibitem{amd}
A.~Ono, H.~Horiuchi, T.~Maruyama and A.~Ohnishi.
\newblock Phys.~Rev.~Lett.~{\bf 68}, 2898 (1992).



\bibitem{ai86}
J.~Aichelin and H.~St\"ocker.
\newblock  Phys.~Lett.~{\bf B176}, 14 (1986).

\bibitem{qmdvsqmd}
Compare eg.
% N. Herrmann et al. Nucl. Phys. A or 
 \cite{ai91,boh91} with 
T. Wienold  et al, submitted to Phys. Rev. Lett. or
\cite{hart,hacorii,varenna}

\bibitem{gossiaux951} P.B. Gossiaux, D. Keane, S. Wang, and J. Aichelin,
Phys. Rev. C {\bf 51}, 3357 (1995).

\bibitem{gossiaux952} P.B. Gossiaux, and J. Aichelin, Phys. Rev. C -submitted.




\bibitem{moli85b}
J.~J~Molitoris and H.~St\"ocker,
\newblock Phys.~Rev~{\bf C32}, R346 (1985).



\bibitem{ue33}
E.~A. Uehling and G.~E. Uhlenbeck.
\newblock  Phys. Rev. {\bf 43}, 552 (1933) and
Phys. Rev. {\bf 44}, 917 (1934).

\bibitem{suraud}
E. Suraud, S. Ayik, M. Belkacem, J. Stryjewski,
Nucl. Phys. A542 (1992) 141.

\bibitem{yar79}
Y.~Yariv and Z.~Frankel.
\newblock Phys.~Rev.~{\bf C20}, 2227 (1979).

\bibitem{cug80}
J.~Cugnon.
\newblock  Phys.~Rev.~{\bf C22}, 1885 (1980).

%**
\bibitem{pei89}
G.~Peilert, H.~St\"ocker, A.~Rosenhauer, A.~Bohnet, J.~Aichelin and W.~Greiner.
\newblock Phys.~Rev.~{\bf C39}, 1402 (1989).


\bibitem{bod77}
A.R.~Bodmer and C.N.~Panos.
\newblock Phys.~Rev.~{\bf C15}, 1342 (1977).

\bibitem{wil78}
L.~Wilets, Y.~Yariv and R.~Chestnut.
\newblock Nucl.~Phys.~{\bf A301}, 359 (1978).

\bibitem{kis83}
S.M.~Kiselew and Y.E.~Pokrovskil.    
\newblock Sov.~Journ.~Nucl.~Phys.~{\bf 38}, 46 (1983).

\bibitem{mol84}
J.J.~Molitoris, J.B.~Hoffer, H.~Kruse and H.~St\"ocker.
\newblock Phys.~Rev.~Lett.~{\bf 53}, 899 (1984). 


\bibitem{KermannKoonin}
A.K. Kermann and S.E.Koonin, Ann. Phys. 100 (1976) 332.

\bibitem{ar82}
L.~G.~Arnold et al.
\newblock Phys.~Rev.~{\bf C25}, 936 (1982).

\bibitem{pa67}
G. Passatore.
\newblock Nucl.~Phys. {\bf A95}, 694 (1967).

\bibitem{bert88b}
G.~F.~Bertsch and S.~Das~Gupta,
\newblock Phys.~Rep.~{\bf 160}, 189 (1988).

\bibitem{ai87b}
J.~Aichelin, A.~Rosenhauer, G.~Peilert, H.~St\"ocker, and W.~Greiner.
\newblock  Phys.~Rev.~Lett.~{\bf 58}, 1926 (1987).

\bibitem{hama90}
S.~Hama et al. 
\newblock Phys.~Rev.~{\bf C41}, 2737 (1990).

\bibitem{newopt}
Ch.~Hartnack and J.~Aichelin
\newblock Phys.~Rev.~{\bf C49}, 2801 (1994).

\bibitem{malfliet} 
W. Botermanns and R.~Malfliedt,
\newblock Phys.~Lett. {\bf B 215}, 617 (1988) and 
\newblock Phys.~Rep.~{\bf 198}, 115 (1990).
 
\bibitem{mosel} 
W.~Cassing and U.~Mosel,
\newblock Prog.~Nucl.~Part,~Phys.{\bf 25}, 1 (1990)

\bibitem{pqmd}
G.~Peilert, J.~Konopka, M.~Blann, M.~G.~Mustafa, H.~St\"ocker and W. Greiner.
\newblock Phys.~Rev.~{\bf C46}, 1457 (1992).

\bibitem{cug81}
J.~Cugnon, T.~Mizutani and J.~Vandermeulen.
\newblock Nucl.~Phys.~{\bf A352}, 505 (1981).

\bibitem{kodama}
T. Kodama et al, Phys. Rev. C29 (1984) 2146 .

\bibitem{hart}
Ch. Hartnack.
\newblock PhD thesis, GSI-Report 93-5 (1993).

\bibitem{ai89}
J.~Aichelin, C.~Hartnack, A.~Bohnet, Li~Zhuxia, G.~Peilert, H.~St\"ocker
and W.~Greiner.
\newblock Phys.~Lett.~{\bf B 224}, 34 (1989). 

\bibitem{boh91}
A.~Bohnet et al. 
\newblock Phys.~Rev.~{\bf C44}, 2111 (1991).

\bibitem{beg93}
M.~Begemann-Blaich et al.
\newblock Phys.~Rev.~{\bf C 48}, 610 (1993).

\bibitem{muell93}
W.F.J.~M\"uller et al.
\newblock Phys. Lett. {\bf B 298}, 27 (1993)

\bibitem{jeong}
S.C. Jeong et al. 
\newblock Phys.~Rev.~Lett. 72, 3468 (1994).

\bibitem{goss94}
P.B.~Gossiaux et al.
\newblock Phys.~Rev. {\bf C 51}, 3357 (1995).

\bibitem{hub94}
S.~Huber and J.~Aichelin.
\newblock Nucl.~Phys.~{\bf A573}, 587 (1994).
 

\bibitem{cug89}
J. Cugnon, private communication 
 
\bibitem{ha89}
C.~Hartnack, L.~Zhuxia, L.~Neise, G.~Peilert, A.~Rosenhauer, H.~Sorge,
  J.~Aichelin, H.~St\"ocker, and W.~Greiner.
\newblock Nucl.~Phys.~{\bf A495}, 303 (1989).


\bibitem{vw82}
B.~J. VerWest and R.~A. Arndt.
\newblock Phys.~Rev.~{\bf C25}, 1979 (1982).

\bibitem{ha88}
Ch. Hartnack, H.~St\"ocker, and W.~Greiner.
\newblock In H.~Feldmeier, editor, {\em Proc. of the International Workshop on
  Gross Properties of Nuclei and Nuclear Excitation, XVI, Hirschegg,
  Kleinwalsertal, Austria} (1988).


\bibitem{baprl}
S.~A.~Bass, C.~Hartnack, H.~St\"ocker and W.~Greiner.
\newblock Phys.~Rev.~Lett.~{\bf 71}, 1144 (1993).

\bibitem{baprc}
S.~A.~Bass, C.~Hartnack, H.~St\"ocker and W.~Greiner.
\newblock Phys.~Rev.~{\bf C 51 }, 3343 (1994).

\bibitem{mpl}
Ch.~Hartnack, J.~Aichelin, H.~St\"ocker and W.~Greiner.
\newblock Mod.~Phys.~Lett.~{\bf A9}, 1151 (1994). 

\bibitem{chplb94}
Ch.~Hartnack, J.~Aichelin, H.~St\"ocker and W.~Greiner.
\newblock Phys.~Lett. {\bf B 336}, 131 (1994).

\bibitem{ssoff}
S.~Soff et al. 
\newblock Phys.~Rev. {\bf C 51}, 3320 (1995)

\bibitem{madey}
R. Madey, W.M.~Zhang et al. 
\newblock Phys.~Rev. {\bf C 42}, 1068 (1990)

\bibitem{claesson}
G.~Claesson et al.
\newblock Phys.~Lett. {\bf B 251}, 23 (1990)  

\bibitem{ramilien}
V.~Ramilien et al. 
\newblock Nucl.~Phys. {\bf A 587}, 802 (1995)

\bibitem{Dorso87}
C.\ Dorso, S.\ Duarte, and J.\ Randrup,
Phys.\ Lett.\ B {\bf 188}, 287 (1987). 


\bibitem{Peilert91}
G.\ Peilert, J.\ Randrup, H.\ St\"ocker, and W.\ Greiner,
Phys.\ Lett.\ B {\bf 260}, 271 (1991).

\bibitem{konopka}
J.\ Konopka, H.\ St\"ocker, and W.\ Greiner,
Nucl.\ Phys.\ {\bf A583}, 357c (1995).
%\bibitem{kuhn93}
%Ch.~Kuhn et al.
%\newblock Phys.~Rev. {\bf C48}, 1232 (1993). 
%%% {\bf oder war da nur QSM drin?}


\bibitem{jutta}
J.~Jaenicke, J.~Aichelin, N. Ohtsuka, R.~Linden, A.~Faessler,
\newblock Nucl.~Phys. {\bf A536}, 201 (1992). 

\bibitem{kaon94}
C. Hartnack, J. Jaenicke, L. Sehn, H. St\"ocker, J. Aichelin,
\newblock Nucl. Phys. {\bf A 580}, 643 (1994). 

%\bibitem{jeba}
%J.~Konopka and S.A.~Bass, private 
%\end{thebibliography}

\bibitem{Sorge}
H. Sorge, H. St\"ocker, W. Greiner, Ann. Phys. 192, 266 (1989). 

\bibitem{leh95}
E.~Lehmann, R.K.~Puri, A.~Faessler, G.~Batko, S.W.~Huang.
\newblock Phys.~Rev.~{\bf c 51}, 2113 (1995).
 
\bibitem{pu95}
R.K.~Puri, E.~Lehmann, A.~Faessler, S.W.~Huang.
\newblock Zeitsch.~f.~Physik {\bf A 351}, 59 (1995).


\bibitem{hacorii}
Ch. Hartnack, S.A.~Bass, J.~Aichelin, H.~St\"ocker and W.~Greiner,
\newblock In J.~Aichelin and D.~Ardouin, editors, {\em Int. Workshop on 
Multiparticle
Correlations and Nuclear Reactions (Corinne II), Nantes}, World Scientific,
Singapore (1994).

%\bibitem{hadipl}
%Ch.~Hartnack, diploma thesis, University of Frankfurt (unpublished)

\bibitem{ha92}
Ch.~Hartnack et al. 
\newblock Nucl.~Phys.~{\bf A 538}, 53 (1992).

\bibitem{dnsiso}
Ch.~Hartnack, J.~Aichelin, H.~St\"ocker and W.~Greiner,
\newblock Phys.~Rev.~Lett. 72, 3767 (1994).   

\bibitem{aijae}
J. Aichelin and J. Jaenicke, Proc. Workshop on Meson production, 
interaction and decay, Cracow (Poland), may 1991, ed. A. Magiera,
World Scientific (Singapore) 1992.

\bibitem{varenna}
Ch.~Hartnack, J.~Aichelin, H.~St\"ocker, W. Greiner,
\newblock Proc. of the 7th Intern. Conf. on Nucl. Reaction Mechanisms,
Varenna (Italy), June 1994, ed. E. Gadioli, Ricerca Scientifica ed 
Educazione Permanente, Univ. di Milano (Italy), p.242.

\bibitem{Lang92}
A.~Lang et al. 
\newblock Nucl.~Phys. A541, 507 (1992). 

\bibitem{RanKo}
J.~Randrup and C.M.~Ko, 
\newblock Nucl.~Phys. A 343, 519 (1980).

\bibitem{hstprivat} 
H. St\"ocker, private communication 

\bibitem{pochodzalla}
J. Pochodzalla et al. Phys. Rev. Lett. 75 (1995) 1048.

\bibitem{botvina}
A.S. Botvina et al., Nucl. Phys. A 584 (1995) 737.

\bibitem{donangelo}
R. Donangelo and S.R. Souza, Phys. Rev.C 52 (1995) 326.

\bibitem{peniscola}
Ch. Hartnack, H. St\"ocker and W. Greiner,
Proc. of the NATO Advanced Study Institute on the Nuclear Equation of State,
Peniscola (Spain), 1989, ed. H. St\"ocker and W. Greiner, Plenum, New York,
NASI series B216 A, p. 239.

\bibitem{uqmd} 
S.A. Bass et al., Proc. on the Int. Conference on modern physics
at the turn of the millenium, Wilderness (South Africa), World
Scientific, ed. A. Gallmann and H. St\"ocker.
\\
L. Winckelmann et al., Contribution to the
proceedings of the International Conference on Quark Matter 1996, 
Heidelberg (Germany).
\end{references}
\end{document}